\documentclass[twocolumn,secnumarabic,amssymb, nobibnotes, aps, prl]{revtex4-2}

\usepackage{color}
\usepackage[pdftex]{graphicx}

\begin{document}

\title{{Shortening X-ray Pulse Duration} via Saturable Absorption}

\author{Ichiro Inoue$^{1}$}%
\email{inoue@spring8.or.jp}
\author{Yuichi Inubushi$^{1,2}$}
\author{Taito Osaka$^{1}$}
\author{Jumpei Yamada$^{1}$}
\author{Kenji Tamasaku$^{1}$}
\author{Hitoki Yoneda$^{3}$}
\author{Makina Yabashi$^{1,2}$}
\affiliation{$^1$RIKEN SPring-8 Center, 1-1-1 Kouto, Sayo, Hyogo 679-5148, Japan.\\$^2$Japan Synchrotron Radiation Research Institute, Kouto 1-1-1, Sayo, Hyogo 679-5198, Japan.\\$^3$University of Electro-Communications, Chofugaoka 1-5-1, Chofu, Tokyo 182-8585, Japan.}
\begin{abstract}
To shorten the duration of x-ray pulses, we present a nonlinear optical technique {using atoms with core-hole vacancies (core-hole atoms) generated by inner-shell photoionization}. The weak Coulomb screening in the core-hole atoms results in decreased absorption at photon energies immediately above the absorption edge. By employing this phenomenon, referred to as saturable absorption, we successfully reduce the duration of x-ray free-electron laser pulses (photon energy: 9.000 keV, duration: 6-7 fs, fluence: 2.0-3.5$\times$10$^5$ J/cm$^2$) by $\sim$35\%. This finding that core-hole atoms are applicable to nonlinear x-ray optics is an essential stepping stone for extending nonlinear technologies commonplace at optical wavelengths to the hard x-ray region.
\end{abstract}

\maketitle

The interaction of light with matter deviates from a linear response with increasing intensity. The nonlinear responses of light properties such as frequency, polarization, and phase, lie at the heart of modern laser technologies.

 The history of nonlinear optics in the hard x-ray region can be traced back to a pioneering experiment by Eisenberger and McCall in 1971 \cite{EisenbergerPRL1971}, in which they generated two 8.5-keV x-ray photons from a single 17-keV photon through parametric down-conversion (PDC) in beryllium. The detailed features of x-ray PDC became accessible to study \cite{YodaJSR1998, AdamsJSR2000, TamasakuPRL2007, TamasakuActaA2007} with the realization of synchrotron sources, leading to new applications such as ghost imaging \cite{SoferPRX2019,KleinOE2019} and the visualization of local optical response at atomic resolution \cite{TamasakuNP2011}. However,  x-ray nonlinear effects that involve interactions of multiple incident photons were elusive until the 2010s, when the dramatic increase in brilliance made avairable by x-ray free-electron lasers (XFELs) \cite{McNeilNP2010} begun to revolutionize the field of x-ray nonlinear  optics. The XFELs now provide the opportunities to access the second- and higher-order x-ray nonlinear processes, including sum-frequency generation \cite{GloverNature2012}, second harmonic generation \cite{ShwartzPRL2014, LiuPRA2019}, non-sequential two-photon absorption \cite{TamasakuNP2014, GhimirePRA2016, TamasakuPRL2018}, and nonlinear Compton scattering \cite{FuchsNP2015}.
  
  These studies were aimed at achieving a fundamental understanding of the nonlinear x-ray response to matter, especially the determination of cross-sections. {Controlling optical properties with the nonlinear interactions, which has been the jurisdiction of optical lasers \cite{Boyd2008} and extreme-ultraviolet sources based on high-harmonic generation \cite{GariepyPRL2014, FleischerNP2014, LambertNC2015, GauthierNC2017}, is still largely untouched in the hard x-ray region.} The only exception has been the demonstration of an atomic inner-shell laser pumped by XFEL pulses \cite{YonedaNature2015}, where x-ray pulses with greatly improved temporal coherence were generated.
  
  {Here, we propose and demonstrate a fundamental x-ray manipulation technique with a nonlinear effect, namely, shortening the pulse duration.} In addition to its significance for improving the qualities of pulses from the present XFELs, this technique can be expected to be essential for forthcoming cavity-based XFELs \cite{HuangPRL2006, KimPRL2008} because the duration of the x-ray pulses from these sources is inherently long ($\sim$100 fs) due to their extremely small bandwidth, which is of the order of 10 meV.
  
  Our technique employs saturable absorption (SA), a well-known nonlinear effect which has been previously observed in various wavelength regions \cite{LewisJACS1941, NaglerNP2009, YonedaOE2009, YonedaNC2014,RackstrawPRL2015,  WuPRL2016, ChenPRL2018} and is characterized by the deterioration of absorption properties in matter at high optical intensity. 
  {Recently, Yoneda et al. \cite{YonedaNC2014} demonstrated SA in the hard x-ray region  
using atoms with core-hole vacancies (core-hole atoms) produced by inner-shell photoionization.}
   They showed that irradiation with an intense 7.1-keV XFEL pulse reduces the absorption coefficient of an iron foil due to the weakened Coulomb screening in the core-hole atoms, which causes shifting and broadening of the absorption edge to higher energy.
    {Shortening of the XFEL pulse duration is achievable by using an absorption medium with an absorption edge at an energy slightly lower than the incident photon energy.} 
Here, the photons at the leading edge of the pulse are largely absorbed in the medium, leading to a shift in the absorption edge due to the enhanced population of core-hole atoms. The remainder of the pulse is transmitted through the medium with reduced absorption, effectively reducing the pulse duration. The subject of this study is thus the experimental verification of this idea for an “ultrafast shutter” using copper (Cu) foil as the saturable absorber.

{To realize SA with core-hole atoms, it is necessary to create a large population of the core-hole states, i.e. the photoionization rate ($R_{ph}$) should be comparable to or larger than the decay rate of the core-hole state \cite{YonedaNC2014}. $R_{ph}$ at  low incident x-ray intensities is given by $R_{ph}=\sigma_{ph}I/(\hbar\omega)$, where $\sigma_{ph}$, $I$, $\hbar\omega$ are the inner-shell photoabsorption cross-section per atom in the cold state, the x-ray intensity, and the photon energy, respectively. Thus, the required intensity for SA can be roughly estimated to be $\hbar\omega/(\sigma_{ph}\tau_{c})$, where $\tau_c$ is the core-hole lifetime.}
 In the case of Cu, the required intensity is estimated to be 2$\times 10^{20}$ W/cm$^2$ using the parameters of $\hbar\omega=$ 9 keV, $\sigma_{ph}=2\times 10^{-20}$ cm$^2$, and $\tau_{c} =0.4$ fs. This level of intensity can be readily achieved by focusing an XFEL pulse to a sub-micrometer spot. However, an XFEL pulse with an intensity far exceeding $\hbar\omega/(\sigma_{ph}\tau_{c})$ makes the absorber transparent over the whole duration of the pulse. To shorten the pulse duration, the x-ray intensity needs to be comparable to the required intensity for SA, so that there is a variation in the transmittance over the duration of the pulse.

   The experiment was performed at beamline 3 \cite{TonoNJP2013, YabashiJSR2015} of SACLA \cite{IshikawaNP2012} using self-seeded XFEL pulses \cite{InoueNP2019}. A silicon (111) double-crystal monochromator was used to select 9.000 keV radiation (wavelength $\lambda=0.1378$ nm) with a 1-eV bandwidth (FWHM).  
{Since the bandwidth was smaller than the width of the absorption edge in the cold Cu, which is $\sim$10 eV,
the transmittance was considered to be fairly constant over the whole photon energy range in the incident pulse.}   
The x-ray pulse was focused to a spot size of 130 (H) × 165 (V) nm$^2$ (FWHM) by a Kirkpatrick-Baez (KB) mirror system \cite{YumotoAS2020}.
 {A 10-$\mu$m-thick Cu foil with a surface roughness of less than 1 $\mu$m was placed at the focus, and continuously translated spatially to expose an undamaged surface to each XFEL pulse.
 The wavefront aberration at the focus caused by the roughness was estimated to be less than $\lambda/10$. According to the Rayleigh's quarter-wavelength rule \cite{BornBook1999}, the saturable absorber was presumed not to change the beam shape at the focus.} 
 The shot-by-shot pulse energy was monitored in front of the mirror by a calibrated intensity monitor \cite{TonoNJP2013}. The fluence of each pulse at the focus was 0.5-3.5$\times10^5$ J/cm$^2$, which corresponds to an intensity on the order of 10$^{19}$ W/cm$^2$.

  First, we measured the transmittance of the Cu foil by monitoring the transmitted intensity with a photodiode. Figure 1 shows the transmittance as a function of the incident x-ray fluence. 
The transmittance increased with the fluence and was higher than that for the Cu foil in the cold state (dotted line in Fig. 1). This monotonic increase indicates that the absorption was not fully saturated and that transmittance varied over the duration of the pulse.

\begin{figure}
\includegraphics[width=6cm]{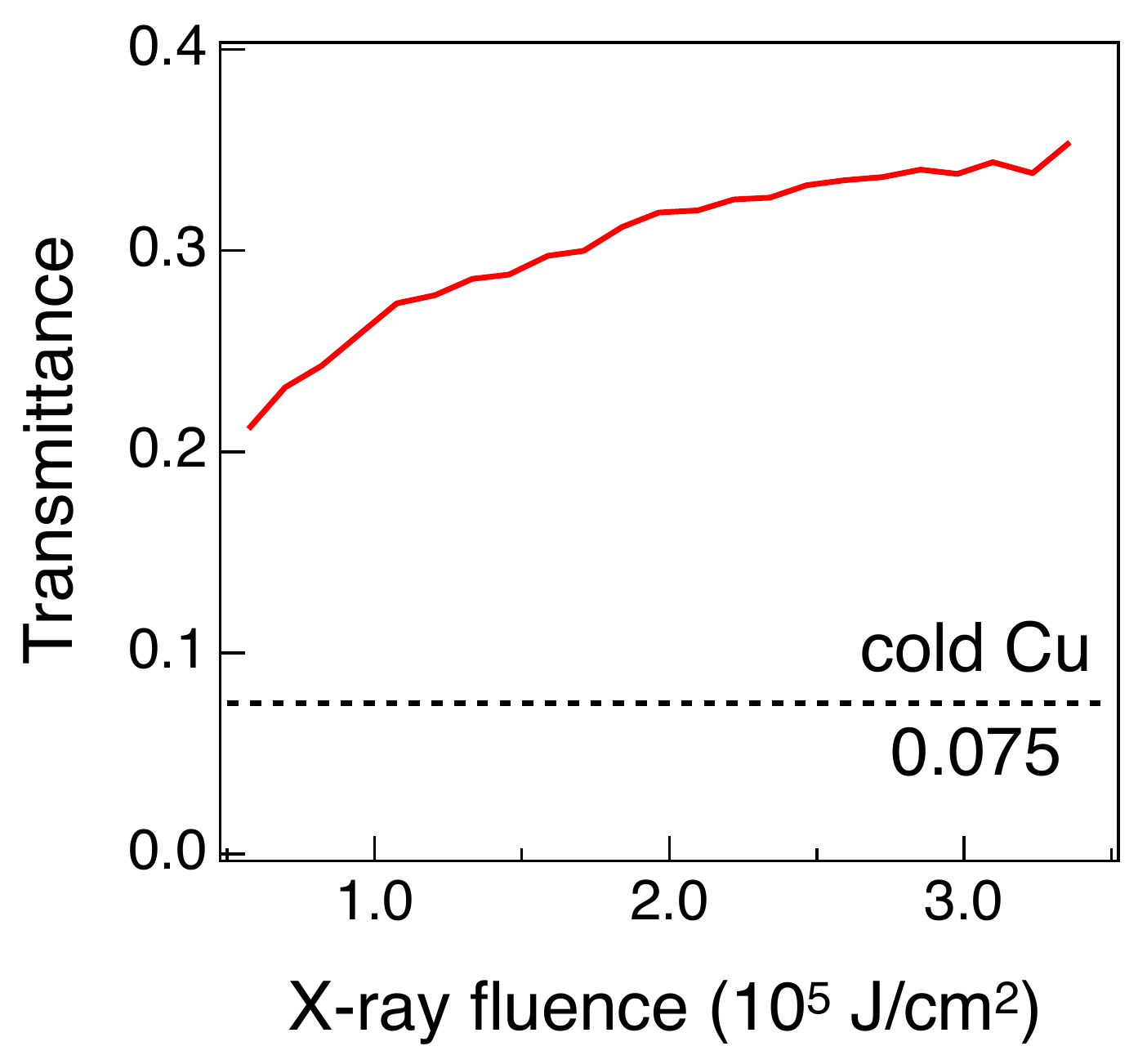}
\caption{Transmittance of the 10-$\mu$m-thick Cu foil at 9.000 keV. The dotted line shows the transmittance of the same foil in the cold state (0.075), which was evaluated by placing the foil at an off-focus position (x-ray fluence of $\sim10^{-4}$ J/cm$^2$)}
\end{figure}

   To evaluate the duration of the x-ray pulse after transmitting through the Cu foil (hereafter called the “transmitted pulse”), we adopted an intensity correlation technique using x-ray fluorescence \cite{InoueJSR2019, InoueJSR2021}. 
   {
   In this technique, the shot-by-shot spatial distribution of the fluorescence from a material irradiated with the x-ray pulse is measured with a two-dimensional detector. Then, the two-point intensity correlation of the fluorescence is evaluated. For pulsed chaotic light, it is known that the degree of intensity correlation of the light is determined by the magnitude relation between the coherence time and the pulse duration\cite{IkonenPRL1992, MiyamotoOL1993, YabashiPRL2002}. Since fluorescence is chaotic, the pulse duration of the fluorescence can be determined from a measurement of the intensity correlation of the fluorescence, and thereby, the x-ray pulse duration can be characterized by taking into account the fluorescence lifetime.}
    
    Figure 2 (a) presents a schematic of the experimental setup. A 20-$\mu$m-thick nickel (Ni) foil was placed 300 $\mu$m downstream from the x-ray focus and used as a source for the fluorescence. The spatial distribution of the Ni $K\alpha$ fluorescence was measured with a multi-port charge-coupled device (MPCCD) detector 
   \cite{KameshimaRSI2014} located 1.6 m downstream of the foil. A 40-$\mu$m-thick cobalt (Co) foil was inserted behind the Ni foil as a filter to reduce contamination from Cu $K\alpha$, Cu $K\beta$ and Ni $K\beta$ fluorescence, and allow imaging only of the Ni $K\alpha$ distribution. Simultaneously, the Ni $K\alpha$ spectrum was measured with a spectrometer in the von Hamos geometry consisting of a silicon (333) curved crystal and another MPCCD detector. Each MPCCD detector was synchronized to the XFEL pulse and recorded $\sim3\times10^5$ images in a shot-by-shot manner. The same measurement without the Cu foil was also performed to measure the duration of the incident XFEL pulse.

\begin{figure}
\includegraphics[width=8.5 cm]{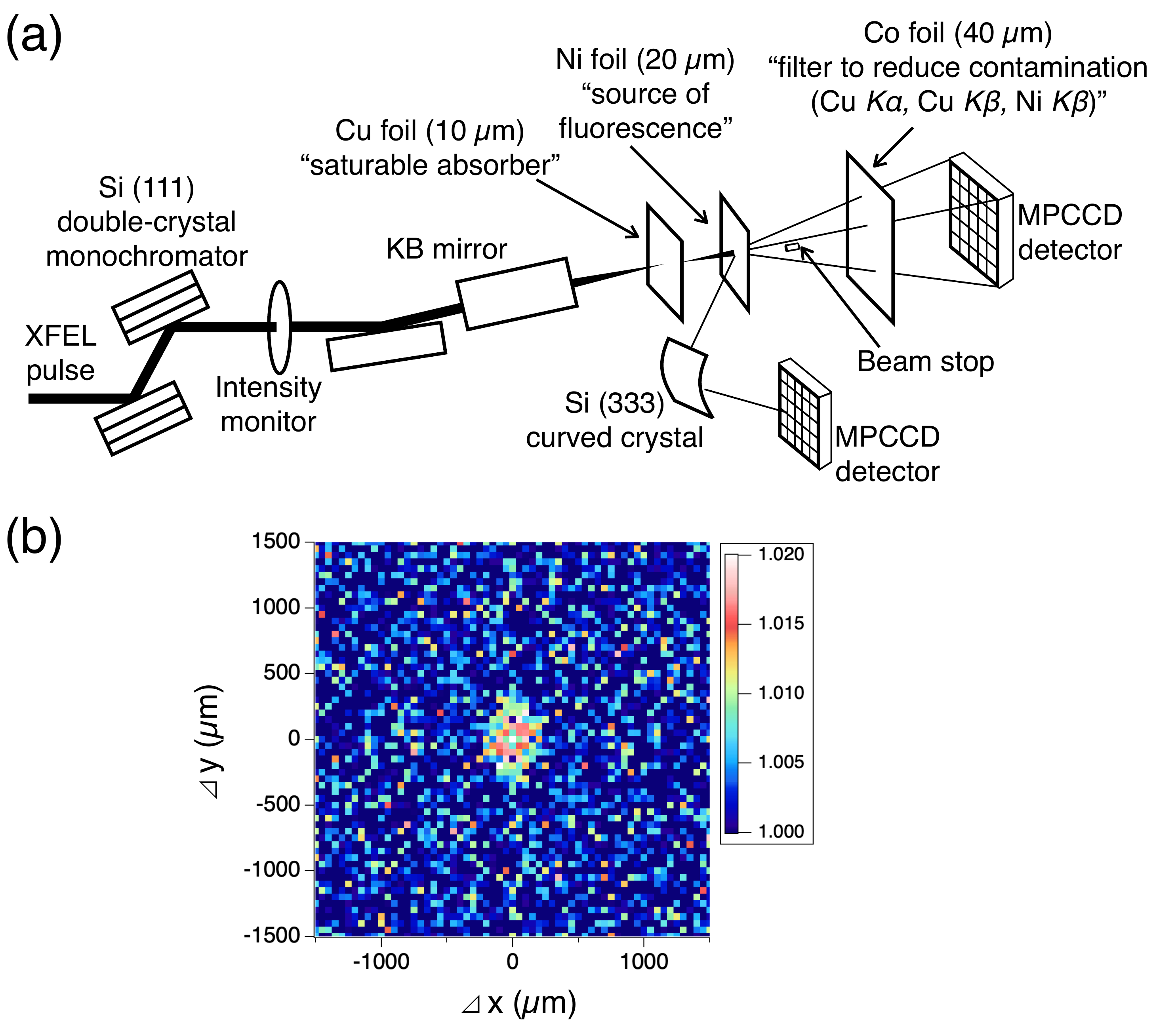}
\caption{(a) Schematic of the experimental setup for evaluating the duration of the x-ray pulse after transmitting through the Cu foil.
(b) Intensity correlation function of the Ni $K\alpha$ fluorescence for an incident x-ray fluence of $(2.56\pm0.14)\times10^5$ J/cm$^2$.
}
\end{figure}

{To evaluate the fluence dependence of the incident and transmitted pulse duration, the fluorescence images were sorted by the fluence at the focus.
By combining the images with the similar fluences, the intensity correlation of the Ni $K\alpha$ fluorescence was evaluated as a function of the horizontal ($\Delta x$) and vertical distances ($\Delta y$) between the two positions by $g_f^{(2)}(\Delta x, \Delta y)=N_0\langle I(x+\Delta x, y+\Delta y)I(x, y)\rangle_{x,y,pulse}$.}
Here, $I(x,y)$ is the intensity of the fluorescence at $(x,y)$, the brackets represent an average over different positions and pulses, and $N_0$ is a normalization factor set to the average value of  $\langle I(x+\Delta x, y+\Delta y)I(x, y)\rangle_{x,y,pulse}$ for large $\Delta x$ and $\Delta y$ ($\sqrt{\Delta x^2+\Delta y^2}>750\ \mu$m). Figure 2 (b) shows a typical example of the measured $g_f^{(2)}(\Delta x, \Delta y)$.
According to \cite{InoueJSR2019, InouePRAB2018}, the intensity correlation function becomes constant ($g_0^{(2)}$) when $\Delta x$ and $\Delta y$ are smaller than the transverse coherence length. $g_0^{(2)}$ is related to the intensity envelope function of the fluorescence ($P(t)$),  the frequency of the fluorescence ($\nu$), and the normalized power spectral density of the fluorescence ($S(\nu)$) as 
\begin{equation}
g_0^{(2)}=1+\frac{1}{2}\int_{-\infty}^{\infty}\Pi(\tau)|\gamma(\tau)|^2d\tau,
\end{equation}
where $\Pi(\tau)=\int_{-\infty}^{\infty}P(t)P(t+\tau)dt/(\int_{-\infty}^{\infty}P(t)dt)^2$ is the normalized autocorrelation function of $P(t)$,  and $\gamma(\tau)=\int_{0}^{\infty} S(\nu)\exp(-i2\pi\nu\tau)d\nu$ is the complex degree of coherence of the fluorescence.

Figure 3 (a) shows $g_0^{(2)}$ of the Ni $K\alpha$ fluorescence for the experiments with and without the Cu foil,  evaluated by fitting the intensity correlation function with a two-dimensional Gaussian function.
{The horizontal error bars represent the range of fluence corresponding to the fluorescence images used for the analysis at each data point,
while the vertical error bars show the uncertainty of $g_0^{(2)}$ from the fitting.}

We assumed that the intensity envelope function for the x-ray pulse impinging on the Ni foil could be described by a Gaussian function. Under this assumption, the intensity envelope function of the Ni $K\alpha$ fluorescence may be expressed as 
\begin{equation}
P(t)=C\exp\left(-\frac{t}{\tau_f}+\frac{\sigma_t^2}{2\tau_{f}^2}\right)\mathrm{erfc}\left(\frac{\sigma_t}{\sqrt{2}\tau_{f}}-\frac{t}{\sqrt{2}\sigma_t}\right),
\end{equation}
where $C$ is a constant of proportionality, $\sigma_t$ is the root-mean-square duration of the x-ray pulse impinging on the Ni foil, $\tau_{f}=0.3$ fs is the lifetime of the Ni $K\alpha$ fluorescence \cite{KrauseJPCRD1979}, and $\mathrm{erfc}$ is the complementary error function, $\mathrm{erfc}(t)=(2/\sqrt{\pi})\int_{t}^{\infty} \exp (-x^2)dx$. 
Since the electron bunch from SACLA has a Gaussian-like current profile \cite{InouePRAB2018}, a Gaussian intensity envelope function for the incident XFEL pulse can be assumed. Regarding the transmitted pulse, the simulation shown below predicts that the intensity envelope function can also be well approximated by a Gaussian function. By substituting eq. (2), $g_0^{(2)}$, and the measured power spectral density of the fluorescence into eq. (1) {(the measured spectrum and the complex degree of coherence of the Ni $K\alpha$ fluorescence are shown in \cite{SupplementalMaterial})}, the duration of the incident and transmitted pulses was evaluated as a function of the x-ray fluence at the focus  [Fig. 3 (b)].
{Here, the errors in the determined pulse duration originate from the uncertainty of $g_0^{(2)}$.}
\begin{figure}
\includegraphics[width=6.5 cm]{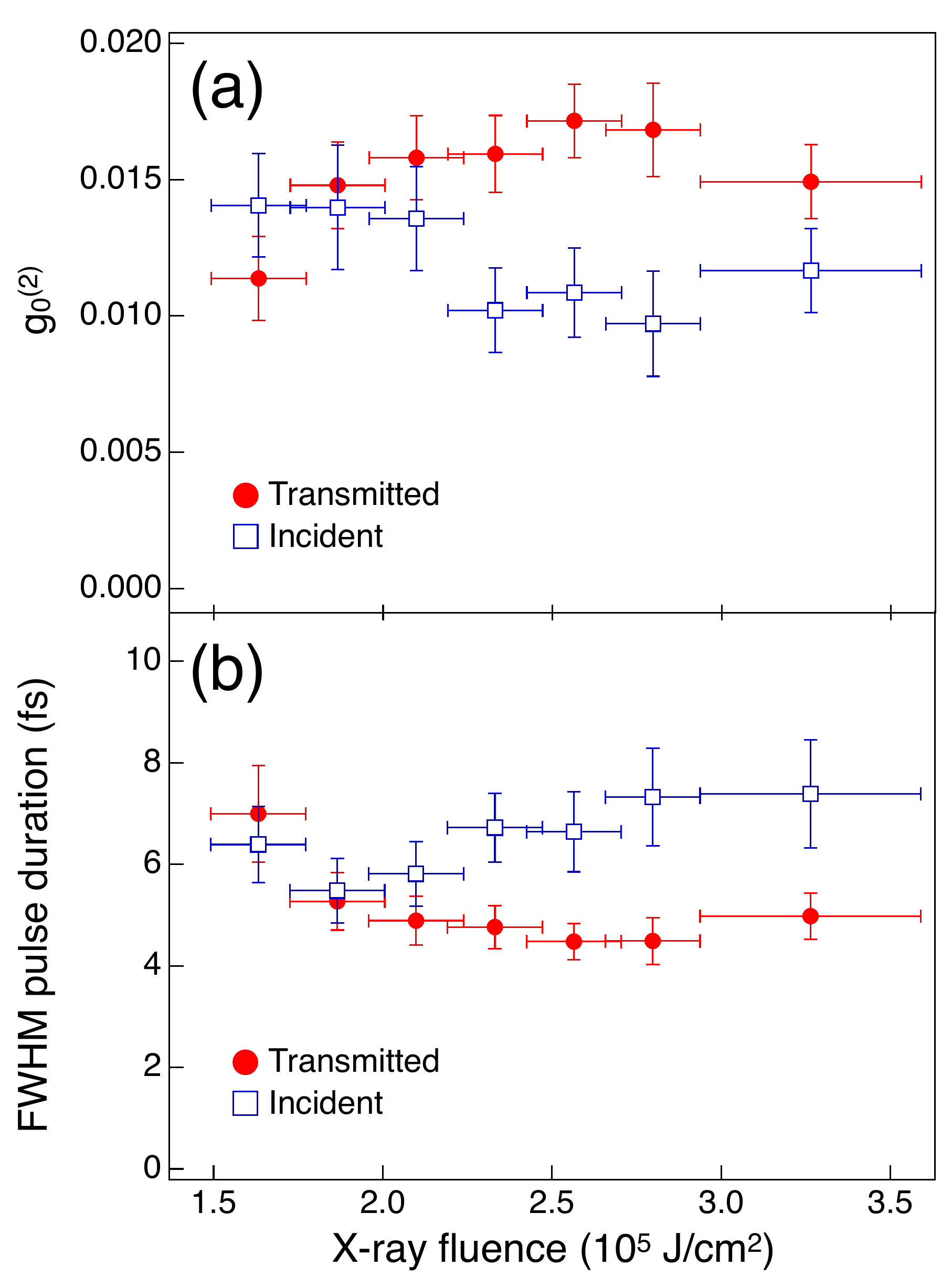}
\caption{(a) $g_0^{(2)}$ of the Ni $K\alpha$ fluorescence for the experiments with (red circles) and without Cu foil (blue squares).
(b) The durations of the incident and transmitted pulses evaluated by the intensity correlation technique.}
\end{figure}
 The incident pulse duration was found to be nearly constant (6-7 fs (FWHM)) regardless of the fluence. For incident pulses with fluences of less than 2.0$\times10^5$ J/cm$^2$, 
the Cu foil did not make a significant difference between the incident and transmitted pulse duration.
For more intense incident pulses (fluences of 2.0-3.5$\times10^5$ J/cm$^2$), the duration of the transmitted pulse was $\sim$4.5 fs (FWHM), i.e., the SA of the Cu foil reduced the pulse duration by $\sim$35\%. Given that the transmittance of the Cu foil was 30-35\% at this fluence [Fig. 1], the peak intensity of the transmitted pulse can be estimated to be as high as half of that of the incident pulse.
  
 { To confirm the validity of the experimental observations and the assumption of the Gaussian-shaped transmitted pulse, we performed numerical calculations of optical propagation through the Cu foil. The simulation was performed for a 7-fs Gaussian incident XFEL pulse with a photon energy of 9.000 keV following the same procedures described in \cite{YonedaNC2014} and solving the rate equations for the populations of the ground and excited states. Here, the rate equations considered photoabsorption, spontaneous and induced emissions of $K\alpha_1$, $K\alpha_2$, and $K\beta$  fluorescence, and $KLL$ and $KMM$ Auger processes. The rate equations were coupled with the optical propagation equation, and the temporal profile of the x-ray pulse at each position inside the Cu foil was calculated.}
      
      Figure 4 (a) shows simulated temporal profiles of the x-ray pulses after transmission through a 10-$\mathrm{\mu m}$-thick Cu foil. The intensity envelope functions for the simulated pulses can be well approximated by Gaussian functions, supporting our earlier assumption made for analyzing the experimental data. Furthermore, the simulation results show good agreement with the fluence dependence of the pulse shortening observed in the experiment; the duration of the transmitted pulse starts to decrease at $\sim$2$\times10^5$ J/cm$^2$, while the duration remains almost constant ($\sim$5 fs) for x-ray fluences of 3-6$\times10^5$ J/cm$^2$.
  [Fig. 4 (b)]. These simulation results support our experimental observations that the SA of the Cu film shortened the x-ray pulse duration.
 {For higher fluences, the duration of the transmitted pulse increases with the fluence and approaches the duration of the incident pulse [Fig. 4 (b)].}
   
\begin{figure}
\includegraphics[width=8cm]{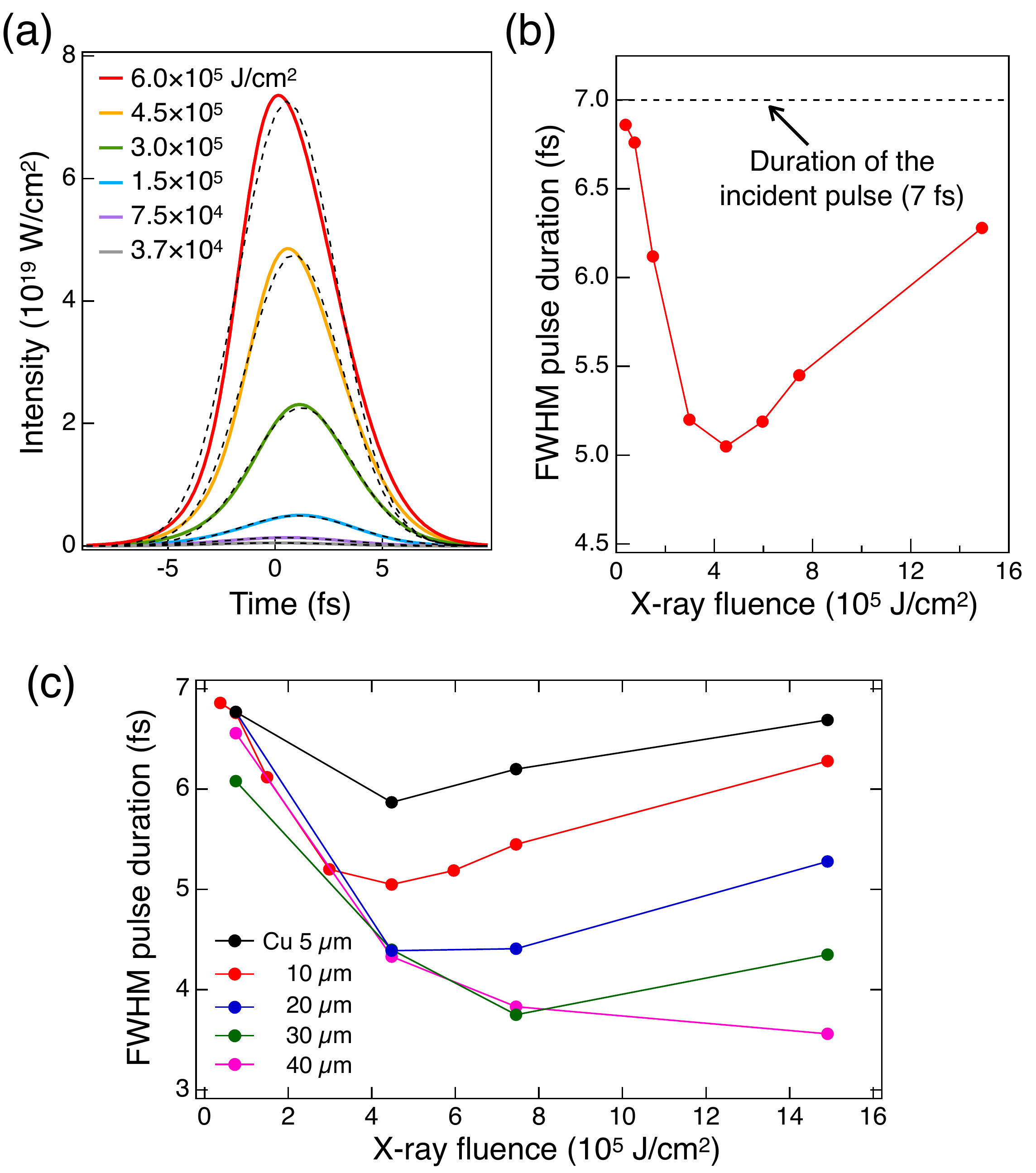}
\caption{(a) Simulated temporal profiles of the x-ray pulses after transmission through a 10-$\mu$m-thick Cu foil. Dotted curves show the Gaussian fits to the simulated profiles.
(b) Duration of the simulated x-ray pulses after transmission through a 10-$\mu$m-thick Cu foil. (c) Duration of the simulated x-ray pulses after transmission through the Cu foils with different thicknesses. }
\end{figure}

{
Finally, we describe the potential applicability of SA for generating shorter x-ray pulses. In principle, the pulse duration can be further shortened by using a thick saturable absorber.
To explicitly show the possibility, we performed numerical calculations of optical propagation through the Cu foils with different thicknesses, following the same procedures described above. 
Figure 4 (c) shows the duration of the transmitted pulse for a 7-fs Gaussian incident XFEL pulse with a photon energy of 9.000 keV. 
Although the required x-ray fluence for reducing the pulse duration increases with the thickness of the foil, one can produce shorter x-ray pulses with a thicker saturable absorber. 
Our previous simulation predicts that the duration of the transmitted pulse can be shortened to the attosecond region by using an x-ray pulse with an intensity on the orders of 10$^{21}$ W/cm$^2$ \cite{YonedaNC2014}.
Notably, such high x-ray intensities have been already realized at SACLA using  state-of-the-art x-ray focusing optics \cite{MatsuyamaSciRep2018, InoueJSR2020}.
The generation and temporal control of attosecond x-ray pulses  by SA are feasible, and fascinating subjects.
Combining the pulse shortening by SA with the recently avariable sub-fs XFEL sources \cite{HuangPRL2017, MarinelliAPL2017} will be promising for realizing single-mode x-ray pulses. }

   {In summary, we demonstrated the shortening of x-ray pulse duration through a nonlinear x-ray interaction with matter.} By using a Cu foil as a saturable absorber, we reduced the duration of a 9.000-keV x-ray pulse by $\sim$35\%. The present study proves the applicability of the core-hole atoms as nonlinear optical devices. This finding should prove to be  an essential step towards the extension of nonlinear technologies commonplace at optical wavelengths to the hard x-ray region.
\\
\acknowledgements{{We acknowledge J. R. Harries for careful reading of the manuscript and  J. St\"{o}hr for pointing out relevant literatures.} The experiments were performed with the approval of the Japan Synchrotron Radiation Research Institute (JASRI, Proposals No. 2019B8012, No. 2020A8021). The work was supported by the Japan Society for the Promotion of Science (JSPS) KAKENHI Grants (No. 19K20604).}

\bibstyle{natbib}
\bibliography{Ref}

\begin{thebibliography}{50}%
\makeatletter
\providecommand \@ifxundefined [1]{%
 \@ifx{#1\undefined}
}%
\providecommand \@ifnum [1]{%
 \ifnum #1\expandafter \@firstoftwo
 \else \expandafter \@secondoftwo
 \fi
}%
\providecommand \@ifx [1]{%
 \ifx #1\expandafter \@firstoftwo
 \else \expandafter \@secondoftwo
 \fi
}%
\providecommand \natexlab [1]{#1}%
\providecommand \enquote  [1]{``#1''}%
\providecommand \bibnamefont  [1]{#1}%
\providecommand \bibfnamefont [1]{#1}%
\providecommand \citenamefont [1]{#1}%
\providecommand \href@noop [0]{\@secondoftwo}%
\providecommand \href [0]{\begingroup \@sanitize@url \@href}%
\providecommand \@href[1]{\@@startlink{#1}\@@href}%
\providecommand \@@href[1]{\endgroup#1\@@endlink}%
\providecommand \@sanitize@url [0]{\catcode `\\12\catcode `\$12\catcode
  `\&12\catcode `\#12\catcode `\^12\catcode `\_12\catcode `\%12\relax}%
\providecommand \@@startlink[1]{}%
\providecommand \@@endlink[0]{}%
\providecommand \url  [0]{\begingroup\@sanitize@url \@url }%
\providecommand \@url [1]{\endgroup\@href {#1}{\urlprefix }}%
\providecommand \urlprefix  [0]{URL }%
\providecommand \Eprint [0]{\href }%
\providecommand \doibase [0]{https://doi.org/}%
\providecommand \selectlanguage [0]{\@gobble}%
\providecommand \bibinfo  [0]{\@secondoftwo}%
\providecommand \bibfield  [0]{\@secondoftwo}%
\providecommand \translation [1]{[#1]}%
\providecommand \BibitemOpen [0]{}%
\providecommand \bibitemStop [0]{}%
\providecommand \bibitemNoStop [0]{.\EOS\space}%
\providecommand \EOS [0]{\spacefactor3000\relax}%
\providecommand \BibitemShut  [1]{\csname bibitem#1\endcsname}%
\let\auto@bib@innerbib\@empty
\bibitem [{\citenamefont {Eisenberger}\ and\ \citenamefont
  {McCall}(1971)}]{EisenbergerPRL1971}%
  \BibitemOpen
  \bibfield  {author} {\bibinfo {author} {\bibfnamefont {P.}~\bibnamefont
  {Eisenberger}}\ and\ \bibinfo {author} {\bibfnamefont {S.~L.}\ \bibnamefont
  {McCall}},\ }\bibfield  {title} {\bibinfo {title} {X-ray parametric
  conversion},\ }\href {https://doi.org/10.1103/PhysRevLett.26.684} {\bibfield
  {journal} {\bibinfo  {journal} {Phys. Rev. Lett.}\ }\textbf {\bibinfo
  {volume} {26}},\ \bibinfo {pages} {684} (\bibinfo {year} {1971})}\BibitemShut
  {NoStop}%
\bibitem [{\citenamefont {Yoda}\ \emph {et~al.}(1998)\citenamefont {Yoda},
  \citenamefont {Suzuki}, \citenamefont {Zhang}, \citenamefont {Hirano},\ and\
  \citenamefont {Kikuta}}]{YodaJSR1998}%
  \BibitemOpen
  \bibfield  {author} {\bibinfo {author} {\bibfnamefont {Y.}~\bibnamefont
  {Yoda}}, \bibinfo {author} {\bibfnamefont {T.}~\bibnamefont {Suzuki}},
  \bibinfo {author} {\bibfnamefont {X.-W.}\ \bibnamefont {Zhang}}, \bibinfo
  {author} {\bibfnamefont {K.}~\bibnamefont {Hirano}},\ and\ \bibinfo {author}
  {\bibfnamefont {S.}~\bibnamefont {Kikuta}},\ }\bibfield  {title} {\bibinfo
  {title} {{X-ray parametric scattering by a diamond crystal}},\ }\href
  {https://doi.org/10.1107/S0909049597020232} {\bibfield  {journal} {\bibinfo
  {journal} {Journal of Synchrotron Radiation}\ }\textbf {\bibinfo {volume}
  {5}},\ \bibinfo {pages} {980} (\bibinfo {year} {1998})}\BibitemShut {NoStop}%
\bibitem [{\citenamefont {Adams}\ \emph {et~al.}(2000)\citenamefont {Adams},
  \citenamefont {Fernandez}, \citenamefont {Lee}, \citenamefont {Materlik},
  \citenamefont {Mills},\ and\ \citenamefont {Novikov}}]{AdamsJSR2000}%
  \BibitemOpen
  \bibfield  {author} {\bibinfo {author} {\bibfnamefont {B.}~\bibnamefont
  {Adams}}, \bibinfo {author} {\bibfnamefont {P.}~\bibnamefont {Fernandez}},
  \bibinfo {author} {\bibfnamefont {W.-K.}\ \bibnamefont {Lee}}, \bibinfo
  {author} {\bibfnamefont {G.}~\bibnamefont {Materlik}}, \bibinfo {author}
  {\bibfnamefont {D.~M.}\ \bibnamefont {Mills}},\ and\ \bibinfo {author}
  {\bibfnamefont {D.~V.}\ \bibnamefont {Novikov}},\ }\bibfield  {title}
  {\bibinfo {title} {{Parametric down conversion of X-ray photons}},\ }\href
  {https://doi.org/10.1107/S0909049599015113} {\bibfield  {journal} {\bibinfo
  {journal} {Journal of Synchrotron Radiation}\ }\textbf {\bibinfo {volume}
  {7}},\ \bibinfo {pages} {81} (\bibinfo {year} {2000})}\BibitemShut {NoStop}%
\bibitem [{\citenamefont {Tamasaku}\ and\ \citenamefont
  {Ishikawa}(2007{\natexlab{a}})}]{TamasakuPRL2007}%
  \BibitemOpen
  \bibfield  {author} {\bibinfo {author} {\bibfnamefont {K.}~\bibnamefont
  {Tamasaku}}\ and\ \bibinfo {author} {\bibfnamefont {T.}~\bibnamefont
  {Ishikawa}},\ }\bibfield  {title} {\bibinfo {title} {Interference between
  compton scattering and x-ray parametric down-conversion},\ }\href
  {https://doi.org/10.1103/PhysRevLett.98.244801} {\bibfield  {journal}
  {\bibinfo  {journal} {Phys. Rev. Lett.}\ }\textbf {\bibinfo {volume} {98}},\
  \bibinfo {pages} {244801} (\bibinfo {year} {2007}{\natexlab{a}})}\BibitemShut
  {NoStop}%
\bibitem [{\citenamefont {Tamasaku}\ and\ \citenamefont
  {Ishikawa}(2007{\natexlab{b}})}]{TamasakuActaA2007}%
  \BibitemOpen
  \bibfield  {author} {\bibinfo {author} {\bibfnamefont {K.}~\bibnamefont
  {Tamasaku}}\ and\ \bibinfo {author} {\bibfnamefont {T.}~\bibnamefont
  {Ishikawa}},\ }\bibfield  {title} {\bibinfo {title} {{Idler energy dependence
  of nonlinear diffraction in X {$\rightarrow$} X + EUV parametric
  down-conversion}},\ }\href {https://doi.org/10.1107/S0108767307032680}
  {\bibfield  {journal} {\bibinfo  {journal} {Acta Crystallographica Section
  A}\ }\textbf {\bibinfo {volume} {63}},\ \bibinfo {pages} {437} (\bibinfo
  {year} {2007}{\natexlab{b}})}\BibitemShut {NoStop}%
\bibitem [{\citenamefont {Sofer}\ \emph {et~al.}(2019)\citenamefont {Sofer},
  \citenamefont {Strizhevsky}, \citenamefont {Schori}, \citenamefont
  {Tamasaku},\ and\ \citenamefont {Shwartz}}]{SoferPRX2019}%
  \BibitemOpen
  \bibfield  {author} {\bibinfo {author} {\bibfnamefont {S.}~\bibnamefont
  {Sofer}}, \bibinfo {author} {\bibfnamefont {E.}~\bibnamefont {Strizhevsky}},
  \bibinfo {author} {\bibfnamefont {A.}~\bibnamefont {Schori}}, \bibinfo
  {author} {\bibfnamefont {K.}~\bibnamefont {Tamasaku}},\ and\ \bibinfo
  {author} {\bibfnamefont {S.}~\bibnamefont {Shwartz}},\ }\bibfield  {title}
  {\bibinfo {title} {Quantum enhanced x-ray detection},\ }\href
  {https://doi.org/10.1103/PhysRevX.9.031033} {\bibfield  {journal} {\bibinfo
  {journal} {Phys. Rev. X}\ }\textbf {\bibinfo {volume} {9}},\ \bibinfo {pages}
  {031033} (\bibinfo {year} {2019})}\BibitemShut {NoStop}%
\bibitem [{\citenamefont {Klein}\ \emph {et~al.}(2019)\citenamefont {Klein},
  \citenamefont {Schori}, \citenamefont {Dolbnya}, \citenamefont {Sawhney},\
  and\ \citenamefont {Shwartz}}]{KleinOE2019}%
  \BibitemOpen
  \bibfield  {author} {\bibinfo {author} {\bibfnamefont {Y.}~\bibnamefont
  {Klein}}, \bibinfo {author} {\bibfnamefont {A.}~\bibnamefont {Schori}},
  \bibinfo {author} {\bibfnamefont {I.~P.}\ \bibnamefont {Dolbnya}}, \bibinfo
  {author} {\bibfnamefont {K.}~\bibnamefont {Sawhney}},\ and\ \bibinfo {author}
  {\bibfnamefont {S.}~\bibnamefont {Shwartz}},\ }\bibfield  {title} {\bibinfo
  {title} {X-ray computational ghost imaging with single-pixel detector},\
  }\href {https://doi.org/10.1364/OE.27.003284} {\bibfield  {journal} {\bibinfo
   {journal} {Opt. Express}\ }\textbf {\bibinfo {volume} {27}},\ \bibinfo
  {pages} {3284} (\bibinfo {year} {2019})}\BibitemShut {NoStop}%
\bibitem [{\citenamefont {Tamasaku}\ \emph {et~al.}(2011)\citenamefont
  {Tamasaku}, \citenamefont {Sawada}, \citenamefont {Nishibori},\ and\
  \citenamefont {Ishikawa}}]{TamasakuNP2011}%
  \BibitemOpen
  \bibfield  {author} {\bibinfo {author} {\bibfnamefont {K.}~\bibnamefont
  {Tamasaku}}, \bibinfo {author} {\bibfnamefont {K.}~\bibnamefont {Sawada}},
  \bibinfo {author} {\bibfnamefont {E.}~\bibnamefont {Nishibori}},\ and\
  \bibinfo {author} {\bibfnamefont {T.}~\bibnamefont {Ishikawa}},\ }\bibfield
  {title} {\bibinfo {title} {Visualizing the local optical response to
  extreme-ultraviolet radiation with a resolution of $\lambda$/380},\ }\href
  {https://doi.org/10.1038/nphys2044} {\bibfield  {journal} {\bibinfo
  {journal} {Nature Physics}\ }\textbf {\bibinfo {volume} {7}},\ \bibinfo
  {pages} {705} (\bibinfo {year} {2011})}\BibitemShut {NoStop}%
\bibitem [{\citenamefont {McNeil}\ and\ \citenamefont
  {Thompson}(2010)}]{McNeilNP2010}%
  \BibitemOpen
  \bibfield  {author} {\bibinfo {author} {\bibfnamefont {B.~W.~J.}\
  \bibnamefont {McNeil}}\ and\ \bibinfo {author} {\bibfnamefont {N.~R.}\
  \bibnamefont {Thompson}},\ }\bibfield  {title} {\bibinfo {title} {X-ray
  free-electron lasers},\ }\href {https://doi.org/10.1038/nphoton.2010.239}
  {\bibfield  {journal} {\bibinfo  {journal} {Nature Photonics}\ }\textbf
  {\bibinfo {volume} {4}},\ \bibinfo {pages} {814} (\bibinfo {year}
  {2010})}\BibitemShut {NoStop}%
\bibitem [{\citenamefont {Glover}\ \emph {et~al.}(2012)\citenamefont {Glover},
  \citenamefont {Fritz}, \citenamefont {Cammarata}, \citenamefont {Allison},
  \citenamefont {Coh}, \citenamefont {Feldkamp}, \citenamefont {Lemke},
  \citenamefont {Zhu}, \citenamefont {Feng}, \citenamefont {Coffee},
  \citenamefont {Fuchs}, \citenamefont {Ghimire}, \citenamefont {Chen},
  \citenamefont {Shwartz}, \citenamefont {Reis}, \citenamefont {Harris},\ and\
  \citenamefont {Hastings}}]{GloverNature2012}%
  \BibitemOpen
  \bibfield  {author} {\bibinfo {author} {\bibfnamefont {T.~E.}\ \bibnamefont
  {Glover}}, \bibinfo {author} {\bibfnamefont {D.~M.}\ \bibnamefont {Fritz}},
  \bibinfo {author} {\bibfnamefont {M.}~\bibnamefont {Cammarata}}, \bibinfo
  {author} {\bibfnamefont {T.~K.}\ \bibnamefont {Allison}}, \bibinfo {author}
  {\bibfnamefont {S.}~\bibnamefont {Coh}}, \bibinfo {author} {\bibfnamefont
  {J.~M.}\ \bibnamefont {Feldkamp}}, \bibinfo {author} {\bibfnamefont
  {H.}~\bibnamefont {Lemke}}, \bibinfo {author} {\bibfnamefont
  {D.}~\bibnamefont {Zhu}}, \bibinfo {author} {\bibfnamefont {Y.}~\bibnamefont
  {Feng}}, \bibinfo {author} {\bibfnamefont {R.~N.}\ \bibnamefont {Coffee}},
  \bibinfo {author} {\bibfnamefont {M.}~\bibnamefont {Fuchs}}, \bibinfo
  {author} {\bibfnamefont {S.}~\bibnamefont {Ghimire}}, \bibinfo {author}
  {\bibfnamefont {J.}~\bibnamefont {Chen}}, \bibinfo {author} {\bibfnamefont
  {S.}~\bibnamefont {Shwartz}}, \bibinfo {author} {\bibfnamefont {D.~A.}\
  \bibnamefont {Reis}}, \bibinfo {author} {\bibfnamefont {S.~E.}\ \bibnamefont
  {Harris}},\ and\ \bibinfo {author} {\bibfnamefont {J.~B.}\ \bibnamefont
  {Hastings}},\ }\bibfield  {title} {\bibinfo {title} {X-ray and optical wave
  mixing},\ }\href {https://doi.org/10.1038/nature11340} {\bibfield  {journal}
  {\bibinfo  {journal} {Nature}\ }\textbf {\bibinfo {volume} {488}},\ \bibinfo
  {pages} {603} (\bibinfo {year} {2012})}\BibitemShut {NoStop}%
\bibitem [{\citenamefont {Shwartz}\ \emph {et~al.}(2014)\citenamefont
  {Shwartz}, \citenamefont {Fuchs}, \citenamefont {Hastings}, \citenamefont
  {Inubushi}, \citenamefont {Ishikawa}, \citenamefont {Katayama}, \citenamefont
  {Reis}, \citenamefont {Sato}, \citenamefont {Tono}, \citenamefont {Yabashi},
  \citenamefont {Yudovich},\ and\ \citenamefont {Harris}}]{ShwartzPRL2014}%
  \BibitemOpen
  \bibfield  {author} {\bibinfo {author} {\bibfnamefont {S.}~\bibnamefont
  {Shwartz}}, \bibinfo {author} {\bibfnamefont {M.}~\bibnamefont {Fuchs}},
  \bibinfo {author} {\bibfnamefont {J.~B.}\ \bibnamefont {Hastings}}, \bibinfo
  {author} {\bibfnamefont {Y.}~\bibnamefont {Inubushi}}, \bibinfo {author}
  {\bibfnamefont {T.}~\bibnamefont {Ishikawa}}, \bibinfo {author}
  {\bibfnamefont {T.}~\bibnamefont {Katayama}}, \bibinfo {author}
  {\bibfnamefont {D.~A.}\ \bibnamefont {Reis}}, \bibinfo {author}
  {\bibfnamefont {T.}~\bibnamefont {Sato}}, \bibinfo {author} {\bibfnamefont
  {K.}~\bibnamefont {Tono}}, \bibinfo {author} {\bibfnamefont {M.}~\bibnamefont
  {Yabashi}}, \bibinfo {author} {\bibfnamefont {S.}~\bibnamefont {Yudovich}},\
  and\ \bibinfo {author} {\bibfnamefont {S.~E.}\ \bibnamefont {Harris}},\
  }\bibfield  {title} {\bibinfo {title} {X-ray second harmonic generation},\
  }\href {https://doi.org/10.1103/PhysRevLett.112.163901} {\bibfield  {journal}
  {\bibinfo  {journal} {Phys. Rev. Lett.}\ }\textbf {\bibinfo {volume} {112}},\
  \bibinfo {pages} {163901} (\bibinfo {year} {2014})}\BibitemShut {NoStop}%
\bibitem [{\citenamefont {Liu}\ \emph {et~al.}(2019)\citenamefont {Liu},
  \citenamefont {Miron}, \citenamefont {\AA{}gren}, \citenamefont {Polyutov},\
  and\ \citenamefont {Gel'mukhanov}}]{LiuPRA2019}%
  \BibitemOpen
  \bibfield  {author} {\bibinfo {author} {\bibfnamefont {J.-C.}\ \bibnamefont
  {Liu}}, \bibinfo {author} {\bibfnamefont {C.}~\bibnamefont {Miron}}, \bibinfo
  {author} {\bibfnamefont {H.}~\bibnamefont {\AA{}gren}}, \bibinfo {author}
  {\bibfnamefont {S.}~\bibnamefont {Polyutov}},\ and\ \bibinfo {author}
  {\bibfnamefont {F.}~\bibnamefont {Gel'mukhanov}},\ }\bibfield  {title}
  {\bibinfo {title} {Resonant x-ray second-harmonic generation in atomic
  gases},\ }\href {https://doi.org/10.1103/PhysRevA.100.063403} {\bibfield
  {journal} {\bibinfo  {journal} {Phys. Rev. A}\ }\textbf {\bibinfo {volume}
  {100}},\ \bibinfo {pages} {063403} (\bibinfo {year} {2019})}\BibitemShut
  {NoStop}%
\bibitem [{\citenamefont {Tamasaku}\ \emph {et~al.}(2014)\citenamefont
  {Tamasaku}, \citenamefont {Shigemasa}, \citenamefont {Inubushi},
  \citenamefont {Katayama}, \citenamefont {Sawada}, \citenamefont {Yumoto},
  \citenamefont {Ohashi}, \citenamefont {Mimura}, \citenamefont {Yabashi},
  \citenamefont {Yamauchi},\ and\ \citenamefont {Ishikawa}}]{TamasakuNP2014}%
  \BibitemOpen
  \bibfield  {author} {\bibinfo {author} {\bibfnamefont {K.}~\bibnamefont
  {Tamasaku}}, \bibinfo {author} {\bibfnamefont {E.}~\bibnamefont {Shigemasa}},
  \bibinfo {author} {\bibfnamefont {Y.}~\bibnamefont {Inubushi}}, \bibinfo
  {author} {\bibfnamefont {T.}~\bibnamefont {Katayama}}, \bibinfo {author}
  {\bibfnamefont {K.}~\bibnamefont {Sawada}}, \bibinfo {author} {\bibfnamefont
  {H.}~\bibnamefont {Yumoto}}, \bibinfo {author} {\bibfnamefont
  {H.}~\bibnamefont {Ohashi}}, \bibinfo {author} {\bibfnamefont
  {H.}~\bibnamefont {Mimura}}, \bibinfo {author} {\bibfnamefont
  {M.}~\bibnamefont {Yabashi}}, \bibinfo {author} {\bibfnamefont
  {K.}~\bibnamefont {Yamauchi}},\ and\ \bibinfo {author} {\bibfnamefont
  {T.}~\bibnamefont {Ishikawa}},\ }\bibfield  {title} {\bibinfo {title} {X-ray
  two-photon absorption competing against single and sequential multiphoton
  processes},\ }\href {https://doi.org/10.1038/nphoton.2014.10} {\bibfield
  {journal} {\bibinfo  {journal} {Nature Photonics}\ }\textbf {\bibinfo
  {volume} {8}},\ \bibinfo {pages} {313} (\bibinfo {year} {2014})}\BibitemShut
  {NoStop}%
\bibitem [{\citenamefont {Ghimire}\ \emph {et~al.}(2016)\citenamefont
  {Ghimire}, \citenamefont {Fuchs}, \citenamefont {Hastings}, \citenamefont
  {Herrmann}, \citenamefont {Inubushi}, \citenamefont {Pines}, \citenamefont
  {Shwartz}, \citenamefont {Yabashi},\ and\ \citenamefont
  {Reis}}]{GhimirePRA2016}%
  \BibitemOpen
  \bibfield  {author} {\bibinfo {author} {\bibfnamefont {S.}~\bibnamefont
  {Ghimire}}, \bibinfo {author} {\bibfnamefont {M.}~\bibnamefont {Fuchs}},
  \bibinfo {author} {\bibfnamefont {J.}~\bibnamefont {Hastings}}, \bibinfo
  {author} {\bibfnamefont {S.~C.}\ \bibnamefont {Herrmann}}, \bibinfo {author}
  {\bibfnamefont {Y.}~\bibnamefont {Inubushi}}, \bibinfo {author}
  {\bibfnamefont {J.}~\bibnamefont {Pines}}, \bibinfo {author} {\bibfnamefont
  {S.}~\bibnamefont {Shwartz}}, \bibinfo {author} {\bibfnamefont
  {M.}~\bibnamefont {Yabashi}},\ and\ \bibinfo {author} {\bibfnamefont {D.~A.}\
  \bibnamefont {Reis}},\ }\bibfield  {title} {\bibinfo {title} {Nonsequential
  two-photon absorption from the $k$ shell in solid zirconium},\ }\href
  {https://doi.org/10.1103/PhysRevA.94.043418} {\bibfield  {journal} {\bibinfo
  {journal} {Phys. Rev. A}\ }\textbf {\bibinfo {volume} {94}},\ \bibinfo
  {pages} {043418} (\bibinfo {year} {2016})}\BibitemShut {NoStop}%
\bibitem [{\citenamefont {Tamasaku}\ \emph {et~al.}(2018)\citenamefont
  {Tamasaku}, \citenamefont {Shigemasa}, \citenamefont {Inubushi},
  \citenamefont {Inoue}, \citenamefont {Osaka}, \citenamefont {Katayama},
  \citenamefont {Yabashi}, \citenamefont {Koide}, \citenamefont {Yokoyama},\
  and\ \citenamefont {Ishikawa}}]{TamasakuPRL2018}%
  \BibitemOpen
  \bibfield  {author} {\bibinfo {author} {\bibfnamefont {K.}~\bibnamefont
  {Tamasaku}}, \bibinfo {author} {\bibfnamefont {E.}~\bibnamefont {Shigemasa}},
  \bibinfo {author} {\bibfnamefont {Y.}~\bibnamefont {Inubushi}}, \bibinfo
  {author} {\bibfnamefont {I.}~\bibnamefont {Inoue}}, \bibinfo {author}
  {\bibfnamefont {T.}~\bibnamefont {Osaka}}, \bibinfo {author} {\bibfnamefont
  {T.}~\bibnamefont {Katayama}}, \bibinfo {author} {\bibfnamefont
  {M.}~\bibnamefont {Yabashi}}, \bibinfo {author} {\bibfnamefont
  {A.}~\bibnamefont {Koide}}, \bibinfo {author} {\bibfnamefont
  {T.}~\bibnamefont {Yokoyama}},\ and\ \bibinfo {author} {\bibfnamefont
  {T.}~\bibnamefont {Ishikawa}},\ }\bibfield  {title} {\bibinfo {title}
  {Nonlinear spectroscopy with x-ray two-photon absorption in metallic
  copper},\ }\href {https://doi.org/10.1103/PhysRevLett.121.083901} {\bibfield
  {journal} {\bibinfo  {journal} {Phys. Rev. Lett.}\ }\textbf {\bibinfo
  {volume} {121}},\ \bibinfo {pages} {083901} (\bibinfo {year}
  {2018})}\BibitemShut {NoStop}%
\bibitem [{\citenamefont {Fuchs}\ \emph {et~al.}(2015)\citenamefont {Fuchs},
  \citenamefont {Trigo}, \citenamefont {Chen}, \citenamefont {Ghimire},
  \citenamefont {Shwartz}, \citenamefont {Kozina}, \citenamefont {Jiang},
  \citenamefont {Henighan}, \citenamefont {Bray}, \citenamefont {Ndabashimiye},
  \citenamefont {Bucksbaum}, \citenamefont {Feng}, \citenamefont {Herrmann},
  \citenamefont {Carini}, \citenamefont {Pines}, \citenamefont {Hart},
  \citenamefont {Kenney}, \citenamefont {Guillet}, \citenamefont {Boutet},
  \citenamefont {Williams}, \citenamefont {Messerschmidt}, \citenamefont
  {Seibert}, \citenamefont {Moeller}, \citenamefont {Hastings},\ and\
  \citenamefont {Reis}}]{FuchsNP2015}%
  \BibitemOpen
  \bibfield  {author} {\bibinfo {author} {\bibfnamefont {M.}~\bibnamefont
  {Fuchs}}, \bibinfo {author} {\bibfnamefont {M.}~\bibnamefont {Trigo}},
  \bibinfo {author} {\bibfnamefont {J.}~\bibnamefont {Chen}}, \bibinfo {author}
  {\bibfnamefont {S.}~\bibnamefont {Ghimire}}, \bibinfo {author} {\bibfnamefont
  {S.}~\bibnamefont {Shwartz}}, \bibinfo {author} {\bibfnamefont
  {M.}~\bibnamefont {Kozina}}, \bibinfo {author} {\bibfnamefont
  {M.}~\bibnamefont {Jiang}}, \bibinfo {author} {\bibfnamefont
  {T.}~\bibnamefont {Henighan}}, \bibinfo {author} {\bibfnamefont
  {C.}~\bibnamefont {Bray}}, \bibinfo {author} {\bibfnamefont {G.}~\bibnamefont
  {Ndabashimiye}}, \bibinfo {author} {\bibfnamefont {P.~H.}\ \bibnamefont
  {Bucksbaum}}, \bibinfo {author} {\bibfnamefont {Y.}~\bibnamefont {Feng}},
  \bibinfo {author} {\bibfnamefont {S.}~\bibnamefont {Herrmann}}, \bibinfo
  {author} {\bibfnamefont {G.~A.}\ \bibnamefont {Carini}}, \bibinfo {author}
  {\bibfnamefont {J.}~\bibnamefont {Pines}}, \bibinfo {author} {\bibfnamefont
  {P.}~\bibnamefont {Hart}}, \bibinfo {author} {\bibfnamefont {C.}~\bibnamefont
  {Kenney}}, \bibinfo {author} {\bibfnamefont {S.}~\bibnamefont {Guillet}},
  \bibinfo {author} {\bibfnamefont {S.}~\bibnamefont {Boutet}}, \bibinfo
  {author} {\bibfnamefont {G.~J.}\ \bibnamefont {Williams}}, \bibinfo {author}
  {\bibfnamefont {M.}~\bibnamefont {Messerschmidt}}, \bibinfo {author}
  {\bibfnamefont {M.~M.}\ \bibnamefont {Seibert}}, \bibinfo {author}
  {\bibfnamefont {S.}~\bibnamefont {Moeller}}, \bibinfo {author} {\bibfnamefont
  {J.~B.}\ \bibnamefont {Hastings}},\ and\ \bibinfo {author} {\bibfnamefont
  {D.~A.}\ \bibnamefont {Reis}},\ }\bibfield  {title} {\bibinfo {title}
  {Anomalous nonlinear x-ray compton scattering},\ }\href
  {https://doi.org/10.1038/nphys3452} {\bibfield  {journal} {\bibinfo
  {journal} {Nature Physics}\ }\textbf {\bibinfo {volume} {11}},\ \bibinfo
  {pages} {964} (\bibinfo {year} {2015})}\BibitemShut {NoStop}%
\bibitem [{\citenamefont {Boyd}(2008)}]{Boyd2008}%
  \BibitemOpen
  \bibfield  {author} {\bibinfo {author} {\bibfnamefont {R.~W.}\ \bibnamefont
  {Boyd}},\ }\href@noop {} {\emph {\bibinfo {title} {Nonlinear Optics}}}\
  (\bibinfo  {publisher} {Academic Press},\ \bibinfo {year} {2008})\BibitemShut
  {NoStop}%
\bibitem [{\citenamefont {Gariepy}\ \emph {et~al.}(2014)\citenamefont
  {Gariepy}, \citenamefont {Leach}, \citenamefont {Kim}, \citenamefont
  {Hammond}, \citenamefont {Frumker}, \citenamefont {Boyd},\ and\ \citenamefont
  {Corkum}}]{GariepyPRL2014}%
  \BibitemOpen
  \bibfield  {author} {\bibinfo {author} {\bibfnamefont {G.}~\bibnamefont
  {Gariepy}}, \bibinfo {author} {\bibfnamefont {J.}~\bibnamefont {Leach}},
  \bibinfo {author} {\bibfnamefont {K.~T.}\ \bibnamefont {Kim}}, \bibinfo
  {author} {\bibfnamefont {T.~J.}\ \bibnamefont {Hammond}}, \bibinfo {author}
  {\bibfnamefont {E.}~\bibnamefont {Frumker}}, \bibinfo {author} {\bibfnamefont
  {R.~W.}\ \bibnamefont {Boyd}},\ and\ \bibinfo {author} {\bibfnamefont
  {P.~B.}\ \bibnamefont {Corkum}},\ }\bibfield  {title} {\bibinfo {title}
  {Creating high-harmonic beams with controlled orbital angular momentum},\
  }\href {https://doi.org/10.1103/PhysRevLett.113.153901} {\bibfield  {journal}
  {\bibinfo  {journal} {Phys. Rev. Lett.}\ }\textbf {\bibinfo {volume} {113}},\
  \bibinfo {pages} {153901} (\bibinfo {year} {2014})}\BibitemShut {NoStop}%
\bibitem [{\citenamefont {Fleischer}\ \emph {et~al.}(2014)\citenamefont
  {Fleischer}, \citenamefont {Kfir}, \citenamefont {Diskin}, \citenamefont
  {Sidorenko},\ and\ \citenamefont {Cohen}}]{FleischerNP2014}%
  \BibitemOpen
  \bibfield  {author} {\bibinfo {author} {\bibfnamefont {A.}~\bibnamefont
  {Fleischer}}, \bibinfo {author} {\bibfnamefont {O.}~\bibnamefont {Kfir}},
  \bibinfo {author} {\bibfnamefont {T.}~\bibnamefont {Diskin}}, \bibinfo
  {author} {\bibfnamefont {P.}~\bibnamefont {Sidorenko}},\ and\ \bibinfo
  {author} {\bibfnamefont {O.}~\bibnamefont {Cohen}},\ }\bibfield  {title}
  {\bibinfo {title} {Spin angular momentum and tunable polarization in
  high-harmonic generation},\ }\href {https://doi.org/10.1038/nphoton.2014.108}
  {\bibfield  {journal} {\bibinfo  {journal} {Nature Photonics}\ }\textbf
  {\bibinfo {volume} {8}},\ \bibinfo {pages} {543} (\bibinfo {year}
  {2014})}\BibitemShut {NoStop}%
\bibitem [{\citenamefont {Lambert}\ \emph {et~al.}(2015)\citenamefont
  {Lambert}, \citenamefont {Vodungbo}, \citenamefont {Gautier}, \citenamefont
  {Mahieu}, \citenamefont {Malka}, \citenamefont {Sebban}, \citenamefont
  {Zeitoun}, \citenamefont {Luning}, \citenamefont {Perron}, \citenamefont
  {Andreev}, \citenamefont {Stremoukhov}, \citenamefont {Ardana-Lamas},
  \citenamefont {Dax}, \citenamefont {Hauri}, \citenamefont {Sardinha},\ and\
  \citenamefont {Fajardo}}]{LambertNC2015}%
  \BibitemOpen
  \bibfield  {author} {\bibinfo {author} {\bibfnamefont {G.}~\bibnamefont
  {Lambert}}, \bibinfo {author} {\bibfnamefont {B.}~\bibnamefont {Vodungbo}},
  \bibinfo {author} {\bibfnamefont {J.}~\bibnamefont {Gautier}}, \bibinfo
  {author} {\bibfnamefont {B.}~\bibnamefont {Mahieu}}, \bibinfo {author}
  {\bibfnamefont {V.}~\bibnamefont {Malka}}, \bibinfo {author} {\bibfnamefont
  {S.}~\bibnamefont {Sebban}}, \bibinfo {author} {\bibfnamefont
  {P.}~\bibnamefont {Zeitoun}}, \bibinfo {author} {\bibfnamefont
  {J.}~\bibnamefont {Luning}}, \bibinfo {author} {\bibfnamefont
  {J.}~\bibnamefont {Perron}}, \bibinfo {author} {\bibfnamefont
  {A.}~\bibnamefont {Andreev}}, \bibinfo {author} {\bibfnamefont
  {S.}~\bibnamefont {Stremoukhov}}, \bibinfo {author} {\bibfnamefont
  {F.}~\bibnamefont {Ardana-Lamas}}, \bibinfo {author} {\bibfnamefont
  {A.}~\bibnamefont {Dax}}, \bibinfo {author} {\bibfnamefont {C.~P.}\
  \bibnamefont {Hauri}}, \bibinfo {author} {\bibfnamefont {A.}~\bibnamefont
  {Sardinha}},\ and\ \bibinfo {author} {\bibfnamefont {M.}~\bibnamefont
  {Fajardo}},\ }\bibfield  {title} {\bibinfo {title} {Towards enabling
  femtosecond helicity-dependent spectroscopy with high-harmonic sources},\
  }\href {https://doi.org/10.1038/ncomms7167} {\bibfield  {journal} {\bibinfo
  {journal} {Nature Communications}\ }\textbf {\bibinfo {volume} {6}},\
  \bibinfo {pages} {6167} (\bibinfo {year} {2015})}\BibitemShut {NoStop}%
\bibitem [{\citenamefont {Gauthier}\ \emph {et~al.}(2017)\citenamefont
  {Gauthier}, \citenamefont {Ribi{\v c}}, \citenamefont {Adhikary},
  \citenamefont {Camper}, \citenamefont {Chappuis}, \citenamefont {Cucini},
  \citenamefont {DiMauro}, \citenamefont {Dovillaire}, \citenamefont
  {Frassetto}, \citenamefont {G{\'e}neaux}, \citenamefont {Miotti},
  \citenamefont {Poletto}, \citenamefont {Ressel}, \citenamefont {Spezzani},
  \citenamefont {Stupar}, \citenamefont {Ruchon},\ and\ \citenamefont
  {De~Ninno}}]{GauthierNC2017}%
  \BibitemOpen
  \bibfield  {author} {\bibinfo {author} {\bibfnamefont {D.}~\bibnamefont
  {Gauthier}}, \bibinfo {author} {\bibfnamefont {P.~R.}\ \bibnamefont {Ribi{\v
  c}}}, \bibinfo {author} {\bibfnamefont {G.}~\bibnamefont {Adhikary}},
  \bibinfo {author} {\bibfnamefont {A.}~\bibnamefont {Camper}}, \bibinfo
  {author} {\bibfnamefont {C.}~\bibnamefont {Chappuis}}, \bibinfo {author}
  {\bibfnamefont {R.}~\bibnamefont {Cucini}}, \bibinfo {author} {\bibfnamefont
  {L.~F.}\ \bibnamefont {DiMauro}}, \bibinfo {author} {\bibfnamefont
  {G.}~\bibnamefont {Dovillaire}}, \bibinfo {author} {\bibfnamefont
  {F.}~\bibnamefont {Frassetto}}, \bibinfo {author} {\bibfnamefont
  {R.}~\bibnamefont {G{\'e}neaux}}, \bibinfo {author} {\bibfnamefont
  {P.}~\bibnamefont {Miotti}}, \bibinfo {author} {\bibfnamefont
  {L.}~\bibnamefont {Poletto}}, \bibinfo {author} {\bibfnamefont
  {B.}~\bibnamefont {Ressel}}, \bibinfo {author} {\bibfnamefont
  {C.}~\bibnamefont {Spezzani}}, \bibinfo {author} {\bibfnamefont
  {M.}~\bibnamefont {Stupar}}, \bibinfo {author} {\bibfnamefont
  {T.}~\bibnamefont {Ruchon}},\ and\ \bibinfo {author} {\bibfnamefont
  {G.}~\bibnamefont {De~Ninno}},\ }\bibfield  {title} {\bibinfo {title}
  {Tunable orbital angular momentum in high-harmonic generation},\ }\href
  {https://doi.org/10.1038/ncomms14971} {\bibfield  {journal} {\bibinfo
  {journal} {Nature Communications}\ }\textbf {\bibinfo {volume} {8}},\
  \bibinfo {pages} {14971} (\bibinfo {year} {2017})}\BibitemShut {NoStop}%
\bibitem [{\citenamefont {Yoneda}\ \emph {et~al.}(2015)\citenamefont {Yoneda},
  \citenamefont {Inubushi}, \citenamefont {Nagamine}, \citenamefont {Michine},
  \citenamefont {Ohashi}, \citenamefont {Yumoto}, \citenamefont {Yamauchi},
  \citenamefont {Mimura}, \citenamefont {Kitamura}, \citenamefont {Katayama},
  \citenamefont {Ishikawa},\ and\ \citenamefont {Yabashi}}]{YonedaNature2015}%
  \BibitemOpen
  \bibfield  {author} {\bibinfo {author} {\bibfnamefont {H.}~\bibnamefont
  {Yoneda}}, \bibinfo {author} {\bibfnamefont {Y.}~\bibnamefont {Inubushi}},
  \bibinfo {author} {\bibfnamefont {K.}~\bibnamefont {Nagamine}}, \bibinfo
  {author} {\bibfnamefont {Y.}~\bibnamefont {Michine}}, \bibinfo {author}
  {\bibfnamefont {H.}~\bibnamefont {Ohashi}}, \bibinfo {author} {\bibfnamefont
  {H.}~\bibnamefont {Yumoto}}, \bibinfo {author} {\bibfnamefont
  {K.}~\bibnamefont {Yamauchi}}, \bibinfo {author} {\bibfnamefont
  {H.}~\bibnamefont {Mimura}}, \bibinfo {author} {\bibfnamefont
  {H.}~\bibnamefont {Kitamura}}, \bibinfo {author} {\bibfnamefont
  {T.}~\bibnamefont {Katayama}}, \bibinfo {author} {\bibfnamefont
  {T.}~\bibnamefont {Ishikawa}},\ and\ \bibinfo {author} {\bibfnamefont
  {M.}~\bibnamefont {Yabashi}},\ }\bibfield  {title} {\bibinfo {title} {Atomic
  inner-shell laser at 1.5-{\aa}ngstr{\"o}m wavelength pumped by an x-ray
  free-electron laser},\ }\href {https://doi.org/10.1038/nature14894}
  {\bibfield  {journal} {\bibinfo  {journal} {Nature}\ }\textbf {\bibinfo
  {volume} {524}},\ \bibinfo {pages} {446} (\bibinfo {year}
  {2015})}\BibitemShut {NoStop}%
\bibitem [{\citenamefont {Huang}\ and\ \citenamefont
  {Ruth}(2006)}]{HuangPRL2006}%
  \BibitemOpen
  \bibfield  {author} {\bibinfo {author} {\bibfnamefont {Z.}~\bibnamefont
  {Huang}}\ and\ \bibinfo {author} {\bibfnamefont {R.~D.}\ \bibnamefont
  {Ruth}},\ }\bibfield  {title} {\bibinfo {title} {Fully coherent x-ray pulses
  from a regenerative-amplifier free-electron laser},\ }\href
  {https://doi.org/10.1103/PhysRevLett.96.144801} {\bibfield  {journal}
  {\bibinfo  {journal} {Phys. Rev. Lett.}\ }\textbf {\bibinfo {volume} {96}},\
  \bibinfo {pages} {144801} (\bibinfo {year} {2006})}\BibitemShut {NoStop}%
\bibitem [{\citenamefont {Kim}\ \emph {et~al.}(2008)\citenamefont {Kim},
  \citenamefont {Shvyd'ko},\ and\ \citenamefont {Reiche}}]{KimPRL2008}%
  \BibitemOpen
  \bibfield  {author} {\bibinfo {author} {\bibfnamefont {K.-J.}\ \bibnamefont
  {Kim}}, \bibinfo {author} {\bibfnamefont {Y.}~\bibnamefont {Shvyd'ko}},\ and\
  \bibinfo {author} {\bibfnamefont {S.}~\bibnamefont {Reiche}},\ }\bibfield
  {title} {\bibinfo {title} {A proposal for an x-ray free-electron laser
  oscillator with an energy-recovery linac},\ }\href
  {https://doi.org/10.1103/PhysRevLett.100.244802} {\bibfield  {journal}
  {\bibinfo  {journal} {Phys. Rev. Lett.}\ }\textbf {\bibinfo {volume} {100}},\
  \bibinfo {pages} {244802} (\bibinfo {year} {2008})}\BibitemShut {NoStop}%
\bibitem [{\citenamefont {Lewis}\ \emph {et~al.}(1941)\citenamefont {Lewis},
  \citenamefont {Lipkin},\ and\ \citenamefont {Magel}}]{LewisJACS1941}%
  \BibitemOpen
  \bibfield  {author} {\bibinfo {author} {\bibfnamefont {G.~N.}\ \bibnamefont
  {Lewis}}, \bibinfo {author} {\bibfnamefont {D.}~\bibnamefont {Lipkin}},\ and\
  \bibinfo {author} {\bibfnamefont {T.~T.}\ \bibnamefont {Magel}},\ }\bibfield
  {title} {\bibinfo {title} {Reversible photochemical processes in rigid media.
  a study of the phosphorescent state},\ }\href
  {https://doi.org/10.1021/ja01856a043} {\bibfield  {journal} {\bibinfo
  {journal} {Journal of the American Chemical Society}\ }\textbf {\bibinfo
  {volume} {63}},\ \bibinfo {pages} {3005} (\bibinfo {year}
  {1941})}\BibitemShut {NoStop}%
\bibitem [{\citenamefont {Nagler}\ \emph {et~al.}(2009)\citenamefont {Nagler},
  \citenamefont {Zastrau}, \citenamefont {F{\"a}ustlin}, \citenamefont {Vinko},
  \citenamefont {Whitcher}, \citenamefont {Nelson}, \citenamefont
  {Sobierajski}, \citenamefont {Krzywinski}, \citenamefont {Chalupsky},
  \citenamefont {Abreu}, \citenamefont {Bajt}, \citenamefont {Bornath},
  \citenamefont {Burian}, \citenamefont {Chapman}, \citenamefont {Cihelka},
  \citenamefont {D{\"o}ppner}, \citenamefont {D{\"u}sterer}, \citenamefont
  {Dzelzainis}, \citenamefont {Fajardo}, \citenamefont {F{\"o}rster},
  \citenamefont {Fortmann}, \citenamefont {Galtier}, \citenamefont {Glenzer},
  \citenamefont {G{\"o}de}, \citenamefont {Gregori}, \citenamefont {Hajkova},
  \citenamefont {Heimann}, \citenamefont {Juha}, \citenamefont {Jurek},
  \citenamefont {Khattak}, \citenamefont {Khorsand}, \citenamefont {Klinger},
  \citenamefont {Kozlova}, \citenamefont {Laarmann}, \citenamefont {Lee},
  \citenamefont {Lee}, \citenamefont {Meiwes-Broer}, \citenamefont {Mercere},
  \citenamefont {Murphy}, \citenamefont {Przystawik}, \citenamefont {Redmer},
  \citenamefont {Reinholz}, \citenamefont {Riley}, \citenamefont {R{\"o}pke},
  \citenamefont {Rosmej}, \citenamefont {Saksl}, \citenamefont {Schott},
  \citenamefont {Thiele}, \citenamefont {Tiggesb{\"a}umker}, \citenamefont
  {Toleikis}, \citenamefont {Tschentscher}, \citenamefont {Uschmann},
  \citenamefont {Vollmer}, \citenamefont {Wark},\ and\ \citenamefont
  {et~al.}}]{NaglerNP2009}%
  \BibitemOpen
  \bibfield  {author} {\bibinfo {author} {\bibfnamefont {B.}~\bibnamefont
  {Nagler}}, \bibinfo {author} {\bibfnamefont {U.}~\bibnamefont {Zastrau}},
  \bibinfo {author} {\bibfnamefont {R.~R.}\ \bibnamefont {F{\"a}ustlin}},
  \bibinfo {author} {\bibfnamefont {S.~M.}\ \bibnamefont {Vinko}}, \bibinfo
  {author} {\bibfnamefont {T.}~\bibnamefont {Whitcher}}, \bibinfo {author}
  {\bibfnamefont {A.~J.}\ \bibnamefont {Nelson}}, \bibinfo {author}
  {\bibfnamefont {R.}~\bibnamefont {Sobierajski}}, \bibinfo {author}
  {\bibfnamefont {J.}~\bibnamefont {Krzywinski}}, \bibinfo {author}
  {\bibfnamefont {J.}~\bibnamefont {Chalupsky}}, \bibinfo {author}
  {\bibfnamefont {E.}~\bibnamefont {Abreu}}, \bibinfo {author} {\bibfnamefont
  {S.}~\bibnamefont {Bajt}}, \bibinfo {author} {\bibfnamefont {T.}~\bibnamefont
  {Bornath}}, \bibinfo {author} {\bibfnamefont {T.}~\bibnamefont {Burian}},
  \bibinfo {author} {\bibfnamefont {H.}~\bibnamefont {Chapman}}, \bibinfo
  {author} {\bibfnamefont {J.}~\bibnamefont {Cihelka}}, \bibinfo {author}
  {\bibfnamefont {T.}~\bibnamefont {D{\"o}ppner}}, \bibinfo {author}
  {\bibfnamefont {S.}~\bibnamefont {D{\"u}sterer}}, \bibinfo {author}
  {\bibfnamefont {T.}~\bibnamefont {Dzelzainis}}, \bibinfo {author}
  {\bibfnamefont {M.}~\bibnamefont {Fajardo}}, \bibinfo {author} {\bibfnamefont
  {E.}~\bibnamefont {F{\"o}rster}}, \bibinfo {author} {\bibfnamefont
  {C.}~\bibnamefont {Fortmann}}, \bibinfo {author} {\bibfnamefont
  {E.}~\bibnamefont {Galtier}}, \bibinfo {author} {\bibfnamefont {S.~H.}\
  \bibnamefont {Glenzer}}, \bibinfo {author} {\bibfnamefont {S.}~\bibnamefont
  {G{\"o}de}}, \bibinfo {author} {\bibfnamefont {G.}~\bibnamefont {Gregori}},
  \bibinfo {author} {\bibfnamefont {V.}~\bibnamefont {Hajkova}}, \bibinfo
  {author} {\bibfnamefont {P.}~\bibnamefont {Heimann}}, \bibinfo {author}
  {\bibfnamefont {L.}~\bibnamefont {Juha}}, \bibinfo {author} {\bibfnamefont
  {M.}~\bibnamefont {Jurek}}, \bibinfo {author} {\bibfnamefont {F.~Y.}\
  \bibnamefont {Khattak}}, \bibinfo {author} {\bibfnamefont {A.~R.}\
  \bibnamefont {Khorsand}}, \bibinfo {author} {\bibfnamefont {D.}~\bibnamefont
  {Klinger}}, \bibinfo {author} {\bibfnamefont {M.}~\bibnamefont {Kozlova}},
  \bibinfo {author} {\bibfnamefont {T.}~\bibnamefont {Laarmann}}, \bibinfo
  {author} {\bibfnamefont {H.~J.}\ \bibnamefont {Lee}}, \bibinfo {author}
  {\bibfnamefont {R.~W.}\ \bibnamefont {Lee}}, \bibinfo {author} {\bibfnamefont
  {K.-H.}\ \bibnamefont {Meiwes-Broer}}, \bibinfo {author} {\bibfnamefont
  {P.}~\bibnamefont {Mercere}}, \bibinfo {author} {\bibfnamefont {W.~J.}\
  \bibnamefont {Murphy}}, \bibinfo {author} {\bibfnamefont {A.}~\bibnamefont
  {Przystawik}}, \bibinfo {author} {\bibfnamefont {R.}~\bibnamefont {Redmer}},
  \bibinfo {author} {\bibfnamefont {H.}~\bibnamefont {Reinholz}}, \bibinfo
  {author} {\bibfnamefont {D.}~\bibnamefont {Riley}}, \bibinfo {author}
  {\bibfnamefont {G.}~\bibnamefont {R{\"o}pke}}, \bibinfo {author}
  {\bibfnamefont {F.}~\bibnamefont {Rosmej}}, \bibinfo {author} {\bibfnamefont
  {K.}~\bibnamefont {Saksl}}, \bibinfo {author} {\bibfnamefont
  {R.}~\bibnamefont {Schott}}, \bibinfo {author} {\bibfnamefont
  {R.}~\bibnamefont {Thiele}}, \bibinfo {author} {\bibfnamefont
  {J.}~\bibnamefont {Tiggesb{\"a}umker}}, \bibinfo {author} {\bibfnamefont
  {S.}~\bibnamefont {Toleikis}}, \bibinfo {author} {\bibfnamefont
  {T.}~\bibnamefont {Tschentscher}}, \bibinfo {author} {\bibfnamefont
  {I.}~\bibnamefont {Uschmann}}, \bibinfo {author} {\bibfnamefont {H.~J.}\
  \bibnamefont {Vollmer}}, \bibinfo {author} {\bibfnamefont {J.~S.}\
  \bibnamefont {Wark}},\ and\ \bibinfo {author} {\bibfnamefont {B.~N.}\
  \bibnamefont {et~al.}},\ }\bibfield  {title} {\bibinfo {title} {Turning solid
  aluminium transparent by intense soft x-ray photoionization},\ }\href
  {https://doi.org/10.1038/nphys1341} {\bibfield  {journal} {\bibinfo
  {journal} {Nature Physics}\ }\textbf {\bibinfo {volume} {5}},\ \bibinfo
  {pages} {693} (\bibinfo {year} {2009})}\BibitemShut {NoStop}%
\bibitem [{\citenamefont {Yoneda}\ \emph {et~al.}(2009)\citenamefont {Yoneda},
  \citenamefont {Inubushi}, \citenamefont {Tanaka}, \citenamefont {Yamaguchi},
  \citenamefont {Sato}, \citenamefont {Morimoto}, \citenamefont {Kumagai},
  \citenamefont {Nagasono}, \citenamefont {Higashiya}, \citenamefont {Yabashi},
  \citenamefont {Ishikawa}, \citenamefont {Ohashi}, \citenamefont {Kimura},
  \citenamefont {Kitamura},\ and\ \citenamefont {Kodama}}]{YonedaOE2009}%
  \BibitemOpen
  \bibfield  {author} {\bibinfo {author} {\bibfnamefont {H.}~\bibnamefont
  {Yoneda}}, \bibinfo {author} {\bibfnamefont {Y.}~\bibnamefont {Inubushi}},
  \bibinfo {author} {\bibfnamefont {T.}~\bibnamefont {Tanaka}}, \bibinfo
  {author} {\bibfnamefont {Y.}~\bibnamefont {Yamaguchi}}, \bibinfo {author}
  {\bibfnamefont {F.}~\bibnamefont {Sato}}, \bibinfo {author} {\bibfnamefont
  {S.}~\bibnamefont {Morimoto}}, \bibinfo {author} {\bibfnamefont
  {T.}~\bibnamefont {Kumagai}}, \bibinfo {author} {\bibfnamefont
  {M.}~\bibnamefont {Nagasono}}, \bibinfo {author} {\bibfnamefont
  {A.}~\bibnamefont {Higashiya}}, \bibinfo {author} {\bibfnamefont
  {M.}~\bibnamefont {Yabashi}}, \bibinfo {author} {\bibfnamefont
  {T.}~\bibnamefont {Ishikawa}}, \bibinfo {author} {\bibfnamefont
  {H.}~\bibnamefont {Ohashi}}, \bibinfo {author} {\bibfnamefont
  {H.}~\bibnamefont {Kimura}}, \bibinfo {author} {\bibfnamefont
  {H.}~\bibnamefont {Kitamura}},\ and\ \bibinfo {author} {\bibfnamefont
  {R.}~\bibnamefont {Kodama}},\ }\bibfield  {title} {\bibinfo {title}
  {Ultra-fast switching of light by absorption saturation in vacuum
  ultra-violet region},\ }\href {https://doi.org/10.1364/OE.17.023443}
  {\bibfield  {journal} {\bibinfo  {journal} {Opt. Express}\ }\textbf {\bibinfo
  {volume} {17}},\ \bibinfo {pages} {23443} (\bibinfo {year}
  {2009})}\BibitemShut {NoStop}%
\bibitem [{\citenamefont {Yoneda}\ \emph {et~al.}(2014)\citenamefont {Yoneda},
  \citenamefont {Inubushi}, \citenamefont {Yabashi}, \citenamefont {Katayama},
  \citenamefont {Ishikawa}, \citenamefont {Ohashi}, \citenamefont {Yumoto},
  \citenamefont {Yamauchi}, \citenamefont {Mimura},\ and\ \citenamefont
  {Kitamura}}]{YonedaNC2014}%
  \BibitemOpen
  \bibfield  {author} {\bibinfo {author} {\bibfnamefont {H.}~\bibnamefont
  {Yoneda}}, \bibinfo {author} {\bibfnamefont {Y.}~\bibnamefont {Inubushi}},
  \bibinfo {author} {\bibfnamefont {M.}~\bibnamefont {Yabashi}}, \bibinfo
  {author} {\bibfnamefont {T.}~\bibnamefont {Katayama}}, \bibinfo {author}
  {\bibfnamefont {T.}~\bibnamefont {Ishikawa}}, \bibinfo {author}
  {\bibfnamefont {H.}~\bibnamefont {Ohashi}}, \bibinfo {author} {\bibfnamefont
  {H.}~\bibnamefont {Yumoto}}, \bibinfo {author} {\bibfnamefont
  {K.}~\bibnamefont {Yamauchi}}, \bibinfo {author} {\bibfnamefont
  {H.}~\bibnamefont {Mimura}},\ and\ \bibinfo {author} {\bibfnamefont
  {H.}~\bibnamefont {Kitamura}},\ }\bibfield  {title} {\bibinfo {title}
  {Saturable absorption of intense hard x-rays in iron},\ }\href
  {https://doi.org/10.1038/ncomms6080} {\bibfield  {journal} {\bibinfo
  {journal} {Nature Communications}\ }\textbf {\bibinfo {volume} {5}},\
  \bibinfo {pages} {5080} (\bibinfo {year} {2014})}\BibitemShut {NoStop}%
\bibitem [{\citenamefont {Rackstraw}\ \emph {et~al.}(2015)\citenamefont
  {Rackstraw}, \citenamefont {Ciricosta}, \citenamefont {Vinko}, \citenamefont
  {Barbrel}, \citenamefont {Burian}, \citenamefont {Chalupsk\'y}, \citenamefont
  {Cho}, \citenamefont {Chung}, \citenamefont {Dakovski}, \citenamefont
  {Engelhorn}, \citenamefont {H\'ajkov\'a}, \citenamefont {Heimann},
  \citenamefont {Holmes}, \citenamefont {Juha}, \citenamefont {Krzywinski},
  \citenamefont {Lee}, \citenamefont {Toleikis}, \citenamefont {Turner},
  \citenamefont {Zastrau},\ and\ \citenamefont {Wark}}]{RackstrawPRL2015}%
  \BibitemOpen
  \bibfield  {author} {\bibinfo {author} {\bibfnamefont {D.~S.}\ \bibnamefont
  {Rackstraw}}, \bibinfo {author} {\bibfnamefont {O.}~\bibnamefont
  {Ciricosta}}, \bibinfo {author} {\bibfnamefont {S.~M.}\ \bibnamefont
  {Vinko}}, \bibinfo {author} {\bibfnamefont {B.}~\bibnamefont {Barbrel}},
  \bibinfo {author} {\bibfnamefont {T.}~\bibnamefont {Burian}}, \bibinfo
  {author} {\bibfnamefont {J.}~\bibnamefont {Chalupsk\'y}}, \bibinfo {author}
  {\bibfnamefont {B.~I.}\ \bibnamefont {Cho}}, \bibinfo {author} {\bibfnamefont
  {H.-K.}\ \bibnamefont {Chung}}, \bibinfo {author} {\bibfnamefont {G.~L.}\
  \bibnamefont {Dakovski}}, \bibinfo {author} {\bibfnamefont {K.}~\bibnamefont
  {Engelhorn}}, \bibinfo {author} {\bibfnamefont {V.}~\bibnamefont
  {H\'ajkov\'a}}, \bibinfo {author} {\bibfnamefont {P.}~\bibnamefont
  {Heimann}}, \bibinfo {author} {\bibfnamefont {M.}~\bibnamefont {Holmes}},
  \bibinfo {author} {\bibfnamefont {L.}~\bibnamefont {Juha}}, \bibinfo {author}
  {\bibfnamefont {J.}~\bibnamefont {Krzywinski}}, \bibinfo {author}
  {\bibfnamefont {R.~W.}\ \bibnamefont {Lee}}, \bibinfo {author} {\bibfnamefont
  {S.}~\bibnamefont {Toleikis}}, \bibinfo {author} {\bibfnamefont {J.~J.}\
  \bibnamefont {Turner}}, \bibinfo {author} {\bibfnamefont {U.}~\bibnamefont
  {Zastrau}},\ and\ \bibinfo {author} {\bibfnamefont {J.~S.}\ \bibnamefont
  {Wark}},\ }\bibfield  {title} {\bibinfo {title} {Saturable absorption of an
  x-ray free-electron-laser heated solid-density aluminum plasma},\ }\href
  {https://doi.org/10.1103/PhysRevLett.114.015003} {\bibfield  {journal}
  {\bibinfo  {journal} {Phys. Rev. Lett.}\ }\textbf {\bibinfo {volume} {114}},\
  \bibinfo {pages} {015003} (\bibinfo {year} {2015})}\BibitemShut {NoStop}%
\bibitem [{\citenamefont {Wu}\ \emph {et~al.}(2016)\citenamefont {Wu},
  \citenamefont {Wang}, \citenamefont {Graves}, \citenamefont {Zhu},
  \citenamefont {Schlotter}, \citenamefont {Turner}, \citenamefont {Hellwig},
  \citenamefont {Chen}, \citenamefont {D\"urr}, \citenamefont {Scherz},\ and\
  \citenamefont {St\"ohr}}]{WuPRL2016}%
  \BibitemOpen
  \bibfield  {author} {\bibinfo {author} {\bibfnamefont {B.}~\bibnamefont
  {Wu}}, \bibinfo {author} {\bibfnamefont {T.}~\bibnamefont {Wang}}, \bibinfo
  {author} {\bibfnamefont {C.~E.}\ \bibnamefont {Graves}}, \bibinfo {author}
  {\bibfnamefont {D.}~\bibnamefont {Zhu}}, \bibinfo {author} {\bibfnamefont
  {W.~F.}\ \bibnamefont {Schlotter}}, \bibinfo {author} {\bibfnamefont {J.~J.}\
  \bibnamefont {Turner}}, \bibinfo {author} {\bibfnamefont {O.}~\bibnamefont
  {Hellwig}}, \bibinfo {author} {\bibfnamefont {Z.}~\bibnamefont {Chen}},
  \bibinfo {author} {\bibfnamefont {H.~A.}\ \bibnamefont {D\"urr}}, \bibinfo
  {author} {\bibfnamefont {A.}~\bibnamefont {Scherz}},\ and\ \bibinfo {author}
  {\bibfnamefont {J.}~\bibnamefont {St\"ohr}},\ }\bibfield  {title} {\bibinfo
  {title} {Elimination of x-ray diffraction through stimulated x-ray
  transmission},\ }\href {https://doi.org/10.1103/PhysRevLett.117.027401}
  {\bibfield  {journal} {\bibinfo  {journal} {Phys. Rev. Lett.}\ }\textbf
  {\bibinfo {volume} {117}},\ \bibinfo {pages} {027401} (\bibinfo {year}
  {2016})}\BibitemShut {NoStop}%
\bibitem [{\citenamefont {Chen}\ \emph {et~al.}(2018)\citenamefont {Chen},
  \citenamefont {Higley}, \citenamefont {Beye}, \citenamefont {Hantschmann},
  \citenamefont {Mehta}, \citenamefont {Hellwig}, \citenamefont {Mitra},
  \citenamefont {Bonetti}, \citenamefont {Bucher}, \citenamefont {Carron},
  \citenamefont {Chase}, \citenamefont {Jal}, \citenamefont {Kukreja},
  \citenamefont {Liu}, \citenamefont {Reid}, \citenamefont {Dakovski},
  \citenamefont {F\"ohlisch}, \citenamefont {Schlotter}, \citenamefont
  {D\"urr},\ and\ \citenamefont {St\"ohr}}]{ChenPRL2018}%
  \BibitemOpen
  \bibfield  {author} {\bibinfo {author} {\bibfnamefont {Z.}~\bibnamefont
  {Chen}}, \bibinfo {author} {\bibfnamefont {D.~J.}\ \bibnamefont {Higley}},
  \bibinfo {author} {\bibfnamefont {M.}~\bibnamefont {Beye}}, \bibinfo {author}
  {\bibfnamefont {M.}~\bibnamefont {Hantschmann}}, \bibinfo {author}
  {\bibfnamefont {V.}~\bibnamefont {Mehta}}, \bibinfo {author} {\bibfnamefont
  {O.}~\bibnamefont {Hellwig}}, \bibinfo {author} {\bibfnamefont
  {A.}~\bibnamefont {Mitra}}, \bibinfo {author} {\bibfnamefont
  {S.}~\bibnamefont {Bonetti}}, \bibinfo {author} {\bibfnamefont
  {M.}~\bibnamefont {Bucher}}, \bibinfo {author} {\bibfnamefont
  {S.}~\bibnamefont {Carron}}, \bibinfo {author} {\bibfnamefont
  {T.}~\bibnamefont {Chase}}, \bibinfo {author} {\bibfnamefont
  {E.}~\bibnamefont {Jal}}, \bibinfo {author} {\bibfnamefont {R.}~\bibnamefont
  {Kukreja}}, \bibinfo {author} {\bibfnamefont {T.}~\bibnamefont {Liu}},
  \bibinfo {author} {\bibfnamefont {A.~H.}\ \bibnamefont {Reid}}, \bibinfo
  {author} {\bibfnamefont {G.~L.}\ \bibnamefont {Dakovski}}, \bibinfo {author}
  {\bibfnamefont {A.}~\bibnamefont {F\"ohlisch}}, \bibinfo {author}
  {\bibfnamefont {W.~F.}\ \bibnamefont {Schlotter}}, \bibinfo {author}
  {\bibfnamefont {H.~A.}\ \bibnamefont {D\"urr}},\ and\ \bibinfo {author}
  {\bibfnamefont {J.}~\bibnamefont {St\"ohr}},\ }\bibfield  {title} {\bibinfo
  {title} {Ultrafast self-induced x-ray transparency and loss of magnetic
  diffraction},\ }\href {https://doi.org/10.1103/PhysRevLett.121.137403}
  {\bibfield  {journal} {\bibinfo  {journal} {Phys. Rev. Lett.}\ }\textbf
  {\bibinfo {volume} {121}},\ \bibinfo {pages} {137403} (\bibinfo {year}
  {2018})}\BibitemShut {NoStop}%
\bibitem [{\citenamefont {Tono}\ \emph {et~al.}(2013)\citenamefont {Tono},
  \citenamefont {Togashi}, \citenamefont {Inubushi}, \citenamefont {Sato},
  \citenamefont {Katayama}, \citenamefont {Ogawa}, \citenamefont {Ohashi},
  \citenamefont {Kimura}, \citenamefont {Takahashi}, \citenamefont {Takeshita},
  \citenamefont {Tomizawa}, \citenamefont {Goto}, \citenamefont {Ishikawa},\
  and\ \citenamefont {Yabashi}}]{TonoNJP2013}%
  \BibitemOpen
  \bibfield  {author} {\bibinfo {author} {\bibfnamefont {K.}~\bibnamefont
  {Tono}}, \bibinfo {author} {\bibfnamefont {T.}~\bibnamefont {Togashi}},
  \bibinfo {author} {\bibfnamefont {Y.}~\bibnamefont {Inubushi}}, \bibinfo
  {author} {\bibfnamefont {T.}~\bibnamefont {Sato}}, \bibinfo {author}
  {\bibfnamefont {T.}~\bibnamefont {Katayama}}, \bibinfo {author}
  {\bibfnamefont {K.}~\bibnamefont {Ogawa}}, \bibinfo {author} {\bibfnamefont
  {H.}~\bibnamefont {Ohashi}}, \bibinfo {author} {\bibfnamefont
  {H.}~\bibnamefont {Kimura}}, \bibinfo {author} {\bibfnamefont
  {S.}~\bibnamefont {Takahashi}}, \bibinfo {author} {\bibfnamefont
  {K.}~\bibnamefont {Takeshita}}, \bibinfo {author} {\bibfnamefont
  {H.}~\bibnamefont {Tomizawa}}, \bibinfo {author} {\bibfnamefont
  {S.}~\bibnamefont {Goto}}, \bibinfo {author} {\bibfnamefont {T.}~\bibnamefont
  {Ishikawa}},\ and\ \bibinfo {author} {\bibfnamefont {M.}~\bibnamefont
  {Yabashi}},\ }\bibfield  {title} {\bibinfo {title} {Beamline, experimental
  stations and photon beam diagnostics for the hard x-ray free electron laser
  of {SACLA}},\ }\href {https://doi.org/10.1088/1367-2630/15/8/083035}
  {\bibfield  {journal} {\bibinfo  {journal} {New Journal of Physics}\ }\textbf
  {\bibinfo {volume} {15}},\ \bibinfo {pages} {083035} (\bibinfo {year}
  {2013})}\BibitemShut {NoStop}%
\bibitem [{\citenamefont {Yabashi}\ \emph {et~al.}(2015)\citenamefont
  {Yabashi}, \citenamefont {Tanaka},\ and\ \citenamefont
  {Ishikawa}}]{YabashiJSR2015}%
  \BibitemOpen
  \bibfield  {author} {\bibinfo {author} {\bibfnamefont {M.}~\bibnamefont
  {Yabashi}}, \bibinfo {author} {\bibfnamefont {H.}~\bibnamefont {Tanaka}},\
  and\ \bibinfo {author} {\bibfnamefont {T.}~\bibnamefont {Ishikawa}},\
  }\bibfield  {title} {\bibinfo {title} {{Overview of the SACLA facility}},\
  }\href {https://doi.org/10.1107/S1600577515004658} {\bibfield  {journal}
  {\bibinfo  {journal} {Journal of Synchrotron Radiation}\ }\textbf {\bibinfo
  {volume} {22}},\ \bibinfo {pages} {477} (\bibinfo {year} {2015})}\BibitemShut
  {NoStop}%
\bibitem [{\citenamefont {Ishikawa}\ \emph {et~al.}(2012)\citenamefont
  {Ishikawa}, \citenamefont {Aoyagi}, \citenamefont {Asaka}, \citenamefont
  {Asano}, \citenamefont {Azumi}, \citenamefont {Bizen}, \citenamefont {Ego},
  \citenamefont {Fukami}, \citenamefont {Fukui}, \citenamefont {Furukawa},
  \citenamefont {Goto}, \citenamefont {Hanaki}, \citenamefont {Hara},
  \citenamefont {Hasegawa}, \citenamefont {Hatsui}, \citenamefont {Higashiya},
  \citenamefont {Hirono}, \citenamefont {Hosoda}, \citenamefont {Ishii},
  \citenamefont {Inagaki}, \citenamefont {Inubushi}, \citenamefont {Itoga},
  \citenamefont {Joti}, \citenamefont {Kago}, \citenamefont {Kameshima},
  \citenamefont {Kimura}, \citenamefont {Kirihara}, \citenamefont {Kiyomichi},
  \citenamefont {Kobayashi}, \citenamefont {Kondo}, \citenamefont {Kudo},
  \citenamefont {Maesaka}, \citenamefont {Mar{\'e}chal}, \citenamefont
  {Masuda}, \citenamefont {Matsubara}, \citenamefont {Matsumoto}, \citenamefont
  {Matsushita}, \citenamefont {Matsui}, \citenamefont {Nagasono}, \citenamefont
  {Nariyama}, \citenamefont {Ohashi}, \citenamefont {Ohata}, \citenamefont
  {Ohshima}, \citenamefont {Ono}, \citenamefont {Otake}, \citenamefont {Saji},
  \citenamefont {Sakurai}, \citenamefont {Sato}, \citenamefont {Sawada},
  \citenamefont {Seike}, \citenamefont {Shirasawa}, \citenamefont {Sugimoto},
  \citenamefont {Suzuki}, \citenamefont {Takahashi}, \citenamefont {Takebe},
  \citenamefont {Takeshita}, \citenamefont {Tamasaku}, \citenamefont {Tanaka},
  \citenamefont {Tanaka}, \citenamefont {Tanaka}, \citenamefont {Togashi},
  \citenamefont {Togawa}, \citenamefont {Tokuhisa}, \citenamefont {Tomizawa},
  \citenamefont {Tono}, \citenamefont {Wu}, \citenamefont {Yabashi},
  \citenamefont {Yamaga}, \citenamefont {Yamashita}, \citenamefont {Yanagida},
  \citenamefont {Zhang}, \citenamefont {Shintake}, \citenamefont {Kitamura},\
  and\ \citenamefont {Kumagai}}]{IshikawaNP2012}%
  \BibitemOpen
  \bibfield  {author} {\bibinfo {author} {\bibfnamefont {T.}~\bibnamefont
  {Ishikawa}}, \bibinfo {author} {\bibfnamefont {H.}~\bibnamefont {Aoyagi}},
  \bibinfo {author} {\bibfnamefont {T.}~\bibnamefont {Asaka}}, \bibinfo
  {author} {\bibfnamefont {Y.}~\bibnamefont {Asano}}, \bibinfo {author}
  {\bibfnamefont {N.}~\bibnamefont {Azumi}}, \bibinfo {author} {\bibfnamefont
  {T.}~\bibnamefont {Bizen}}, \bibinfo {author} {\bibfnamefont
  {H.}~\bibnamefont {Ego}}, \bibinfo {author} {\bibfnamefont {K.}~\bibnamefont
  {Fukami}}, \bibinfo {author} {\bibfnamefont {T.}~\bibnamefont {Fukui}},
  \bibinfo {author} {\bibfnamefont {Y.}~\bibnamefont {Furukawa}}, \bibinfo
  {author} {\bibfnamefont {S.}~\bibnamefont {Goto}}, \bibinfo {author}
  {\bibfnamefont {H.}~\bibnamefont {Hanaki}}, \bibinfo {author} {\bibfnamefont
  {T.}~\bibnamefont {Hara}}, \bibinfo {author} {\bibfnamefont {T.}~\bibnamefont
  {Hasegawa}}, \bibinfo {author} {\bibfnamefont {T.}~\bibnamefont {Hatsui}},
  \bibinfo {author} {\bibfnamefont {A.}~\bibnamefont {Higashiya}}, \bibinfo
  {author} {\bibfnamefont {T.}~\bibnamefont {Hirono}}, \bibinfo {author}
  {\bibfnamefont {N.}~\bibnamefont {Hosoda}}, \bibinfo {author} {\bibfnamefont
  {M.}~\bibnamefont {Ishii}}, \bibinfo {author} {\bibfnamefont
  {T.}~\bibnamefont {Inagaki}}, \bibinfo {author} {\bibfnamefont
  {Y.}~\bibnamefont {Inubushi}}, \bibinfo {author} {\bibfnamefont
  {T.}~\bibnamefont {Itoga}}, \bibinfo {author} {\bibfnamefont
  {Y.}~\bibnamefont {Joti}}, \bibinfo {author} {\bibfnamefont {M.}~\bibnamefont
  {Kago}}, \bibinfo {author} {\bibfnamefont {T.}~\bibnamefont {Kameshima}},
  \bibinfo {author} {\bibfnamefont {H.}~\bibnamefont {Kimura}}, \bibinfo
  {author} {\bibfnamefont {Y.}~\bibnamefont {Kirihara}}, \bibinfo {author}
  {\bibfnamefont {A.}~\bibnamefont {Kiyomichi}}, \bibinfo {author}
  {\bibfnamefont {T.}~\bibnamefont {Kobayashi}}, \bibinfo {author}
  {\bibfnamefont {C.}~\bibnamefont {Kondo}}, \bibinfo {author} {\bibfnamefont
  {T.}~\bibnamefont {Kudo}}, \bibinfo {author} {\bibfnamefont {H.}~\bibnamefont
  {Maesaka}}, \bibinfo {author} {\bibfnamefont {X.~M.}\ \bibnamefont
  {Mar{\'e}chal}}, \bibinfo {author} {\bibfnamefont {T.}~\bibnamefont
  {Masuda}}, \bibinfo {author} {\bibfnamefont {S.}~\bibnamefont {Matsubara}},
  \bibinfo {author} {\bibfnamefont {T.}~\bibnamefont {Matsumoto}}, \bibinfo
  {author} {\bibfnamefont {T.}~\bibnamefont {Matsushita}}, \bibinfo {author}
  {\bibfnamefont {S.}~\bibnamefont {Matsui}}, \bibinfo {author} {\bibfnamefont
  {M.}~\bibnamefont {Nagasono}}, \bibinfo {author} {\bibfnamefont
  {N.}~\bibnamefont {Nariyama}}, \bibinfo {author} {\bibfnamefont
  {H.}~\bibnamefont {Ohashi}}, \bibinfo {author} {\bibfnamefont
  {T.}~\bibnamefont {Ohata}}, \bibinfo {author} {\bibfnamefont
  {T.}~\bibnamefont {Ohshima}}, \bibinfo {author} {\bibfnamefont
  {S.}~\bibnamefont {Ono}}, \bibinfo {author} {\bibfnamefont {Y.}~\bibnamefont
  {Otake}}, \bibinfo {author} {\bibfnamefont {C.}~\bibnamefont {Saji}},
  \bibinfo {author} {\bibfnamefont {T.}~\bibnamefont {Sakurai}}, \bibinfo
  {author} {\bibfnamefont {T.}~\bibnamefont {Sato}}, \bibinfo {author}
  {\bibfnamefont {K.}~\bibnamefont {Sawada}}, \bibinfo {author} {\bibfnamefont
  {T.}~\bibnamefont {Seike}}, \bibinfo {author} {\bibfnamefont
  {K.}~\bibnamefont {Shirasawa}}, \bibinfo {author} {\bibfnamefont
  {T.}~\bibnamefont {Sugimoto}}, \bibinfo {author} {\bibfnamefont
  {S.}~\bibnamefont {Suzuki}}, \bibinfo {author} {\bibfnamefont
  {S.}~\bibnamefont {Takahashi}}, \bibinfo {author} {\bibfnamefont
  {H.}~\bibnamefont {Takebe}}, \bibinfo {author} {\bibfnamefont
  {K.}~\bibnamefont {Takeshita}}, \bibinfo {author} {\bibfnamefont
  {K.}~\bibnamefont {Tamasaku}}, \bibinfo {author} {\bibfnamefont
  {H.}~\bibnamefont {Tanaka}}, \bibinfo {author} {\bibfnamefont
  {R.}~\bibnamefont {Tanaka}}, \bibinfo {author} {\bibfnamefont
  {T.}~\bibnamefont {Tanaka}}, \bibinfo {author} {\bibfnamefont
  {T.}~\bibnamefont {Togashi}}, \bibinfo {author} {\bibfnamefont
  {K.}~\bibnamefont {Togawa}}, \bibinfo {author} {\bibfnamefont
  {A.}~\bibnamefont {Tokuhisa}}, \bibinfo {author} {\bibfnamefont
  {H.}~\bibnamefont {Tomizawa}}, \bibinfo {author} {\bibfnamefont
  {K.}~\bibnamefont {Tono}}, \bibinfo {author} {\bibfnamefont {S.}~\bibnamefont
  {Wu}}, \bibinfo {author} {\bibfnamefont {M.}~\bibnamefont {Yabashi}},
  \bibinfo {author} {\bibfnamefont {M.}~\bibnamefont {Yamaga}}, \bibinfo
  {author} {\bibfnamefont {A.}~\bibnamefont {Yamashita}}, \bibinfo {author}
  {\bibfnamefont {K.}~\bibnamefont {Yanagida}}, \bibinfo {author}
  {\bibfnamefont {C.}~\bibnamefont {Zhang}}, \bibinfo {author} {\bibfnamefont
  {T.}~\bibnamefont {Shintake}}, \bibinfo {author} {\bibfnamefont
  {H.}~\bibnamefont {Kitamura}},\ and\ \bibinfo {author} {\bibfnamefont
  {N.}~\bibnamefont {Kumagai}},\ }\bibfield  {title} {\bibinfo {title} {A
  compact x-ray free-electron laser emitting in the sub-{\aa}ngstr{\"o}m
  region},\ }\href {https://doi.org/10.1038/nphoton.2012.141} {\bibfield
  {journal} {\bibinfo  {journal} {Nature Photonics}\ }\textbf {\bibinfo
  {volume} {6}},\ \bibinfo {pages} {540} (\bibinfo {year} {2012})}\BibitemShut
  {NoStop}%
\bibitem [{\citenamefont {Inoue}\ \emph
  {et~al.}(2019{\natexlab{a}})\citenamefont {Inoue}, \citenamefont {Osaka},
  \citenamefont {Hara}, \citenamefont {Tanaka}, \citenamefont {Inagaki},
  \citenamefont {Fukui}, \citenamefont {Goto}, \citenamefont {Inubushi},
  \citenamefont {Kimura}, \citenamefont {Kinjo}, \citenamefont {Ohashi},
  \citenamefont {Togawa}, \citenamefont {Tono}, \citenamefont {Yamaga},
  \citenamefont {Tanaka}, \citenamefont {Ishikawa},\ and\ \citenamefont
  {Yabashi}}]{InoueNP2019}%
  \BibitemOpen
  \bibfield  {author} {\bibinfo {author} {\bibfnamefont {I.}~\bibnamefont
  {Inoue}}, \bibinfo {author} {\bibfnamefont {T.}~\bibnamefont {Osaka}},
  \bibinfo {author} {\bibfnamefont {T.}~\bibnamefont {Hara}}, \bibinfo {author}
  {\bibfnamefont {T.}~\bibnamefont {Tanaka}}, \bibinfo {author} {\bibfnamefont
  {T.}~\bibnamefont {Inagaki}}, \bibinfo {author} {\bibfnamefont
  {T.}~\bibnamefont {Fukui}}, \bibinfo {author} {\bibfnamefont
  {S.}~\bibnamefont {Goto}}, \bibinfo {author} {\bibfnamefont {Y.}~\bibnamefont
  {Inubushi}}, \bibinfo {author} {\bibfnamefont {H.}~\bibnamefont {Kimura}},
  \bibinfo {author} {\bibfnamefont {R.}~\bibnamefont {Kinjo}}, \bibinfo
  {author} {\bibfnamefont {H.}~\bibnamefont {Ohashi}}, \bibinfo {author}
  {\bibfnamefont {K.}~\bibnamefont {Togawa}}, \bibinfo {author} {\bibfnamefont
  {K.}~\bibnamefont {Tono}}, \bibinfo {author} {\bibfnamefont {M.}~\bibnamefont
  {Yamaga}}, \bibinfo {author} {\bibfnamefont {H.}~\bibnamefont {Tanaka}},
  \bibinfo {author} {\bibfnamefont {T.}~\bibnamefont {Ishikawa}},\ and\
  \bibinfo {author} {\bibfnamefont {M.}~\bibnamefont {Yabashi}},\ }\bibfield
  {title} {\bibinfo {title} {Generation of narrow-band x-ray free-electron
  laser via reflection self-seeding},\ }\href
  {https://doi.org/10.1038/s41566-019-0365-y} {\bibfield  {journal} {\bibinfo
  {journal} {Nature Photonics}\ }\textbf {\bibinfo {volume} {13}},\ \bibinfo
  {pages} {319} (\bibinfo {year} {2019}{\natexlab{a}})}\BibitemShut {NoStop}%
\bibitem [{\citenamefont {Yumoto}\ \emph {et~al.}(2020)\citenamefont {Yumoto},
  \citenamefont {Inubushi}, \citenamefont {Osaka}, \citenamefont {Inoue},
  \citenamefont {Koyama}, \citenamefont {Tono}, \citenamefont {Yabashi},\ and\
  \citenamefont {Ohashi}}]{YumotoAS2020}%
  \BibitemOpen
  \bibfield  {author} {\bibinfo {author} {\bibfnamefont {H.}~\bibnamefont
  {Yumoto}}, \bibinfo {author} {\bibfnamefont {Y.}~\bibnamefont {Inubushi}},
  \bibinfo {author} {\bibfnamefont {T.}~\bibnamefont {Osaka}}, \bibinfo
  {author} {\bibfnamefont {I.}~\bibnamefont {Inoue}}, \bibinfo {author}
  {\bibfnamefont {T.}~\bibnamefont {Koyama}}, \bibinfo {author} {\bibfnamefont
  {K.}~\bibnamefont {Tono}}, \bibinfo {author} {\bibfnamefont {M.}~\bibnamefont
  {Yabashi}},\ and\ \bibinfo {author} {\bibfnamefont {H.}~\bibnamefont
  {Ohashi}},\ }\bibfield  {title} {\bibinfo {title} {Nanofocusing optics for an
  x-ray free-electron laser generating an extreme intensity of 100 ew/cm2 using
  total reflection mirrors},\ }\bibfield  {journal} {\bibinfo  {journal}
  {Applied Sciences}\ }\textbf {\bibinfo {volume} {10}},\ \href
  {https://doi.org/10.3390/app10072611} {10.3390/app10072611} (\bibinfo {year}
  {2020})\BibitemShut {NoStop}%
\bibitem [{\citenamefont {Born}\ and\ \citenamefont
  {Wolf}(1999)}]{BornBook1999}%
  \BibitemOpen
  \bibfield  {author} {\bibinfo {author} {\bibfnamefont {M.}~\bibnamefont
  {Born}}\ and\ \bibinfo {author} {\bibfnamefont {E.}~\bibnamefont {Wolf}},\
  }\href@noop {} {\emph {\bibinfo {title} {Principle of Optics}}}\ (\bibinfo
  {publisher} {Cambridge University Press},\ \bibinfo {year}
  {1999})\BibitemShut {NoStop}%
\bibitem [{\citenamefont {Inoue}\ \emph
  {et~al.}(2019{\natexlab{b}})\citenamefont {Inoue}, \citenamefont {Tamasaku},
  \citenamefont {Osaka}, \citenamefont {Inubushi},\ and\ \citenamefont
  {Yabashi}}]{InoueJSR2019}%
  \BibitemOpen
  \bibfield  {author} {\bibinfo {author} {\bibfnamefont {I.}~\bibnamefont
  {Inoue}}, \bibinfo {author} {\bibfnamefont {K.}~\bibnamefont {Tamasaku}},
  \bibinfo {author} {\bibfnamefont {T.}~\bibnamefont {Osaka}}, \bibinfo
  {author} {\bibfnamefont {Y.}~\bibnamefont {Inubushi}},\ and\ \bibinfo
  {author} {\bibfnamefont {M.}~\bibnamefont {Yabashi}},\ }\bibfield  {title}
  {\bibinfo {title} {{Determination of X-ray pulse duration via intensity
  correlation measurements of X-ray fluorescence}},\ }\href
  {https://doi.org/10.1107/S1600577519011202} {\bibfield  {journal} {\bibinfo
  {journal} {Journal of Synchrotron Radiation}\ }\textbf {\bibinfo {volume}
  {26}},\ \bibinfo {pages} {2050} (\bibinfo {year}
  {2019}{\natexlab{b}})}\BibitemShut {NoStop}%
\bibitem [{\citenamefont {Inoue}\ \emph {et~al.}(2021)\citenamefont {Inoue},
  \citenamefont {Tamasaku}, \citenamefont {Osaka}, \citenamefont {Inubushi},\
  and\ \citenamefont {Yabashi}}]{InoueJSR2021}%
  \BibitemOpen
  \bibfield  {author} {\bibinfo {author} {\bibfnamefont {I.}~\bibnamefont
  {Inoue}}, \bibinfo {author} {\bibfnamefont {K.}~\bibnamefont {Tamasaku}},
  \bibinfo {author} {\bibfnamefont {T.}~\bibnamefont {Osaka}}, \bibinfo
  {author} {\bibfnamefont {Y.}~\bibnamefont {Inubushi}},\ and\ \bibinfo
  {author} {\bibfnamefont {M.}~\bibnamefont {Yabashi}},\ }\bibfield  {title}
  {\bibinfo {title} {{Determination of X-ray pulse duration via intensity
  correlation measurement of X-ray fluorescence. Erratum}},\ }\href
  {https://doi.org/10.1107/S1600577520015143} {\bibfield  {journal} {\bibinfo
  {journal} {Journal of Synchrotron Radiation}\ }\textbf {\bibinfo {volume}
  {28}},\ \bibinfo {pages} {372} (\bibinfo {year} {2021})}\BibitemShut
  {NoStop}%
\bibitem [{\citenamefont {Ikonen}(1992)}]{IkonenPRL1992}%
  \BibitemOpen
  \bibfield  {author} {\bibinfo {author} {\bibfnamefont {E.}~\bibnamefont
  {Ikonen}},\ }\bibfield  {title} {\bibinfo {title} {Interference effects
  between independent gamma rays},\ }\href
  {https://doi.org/10.1103/PhysRevLett.68.2759} {\bibfield  {journal} {\bibinfo
   {journal} {Phys. Rev. Lett.}\ }\textbf {\bibinfo {volume} {68}},\ \bibinfo
  {pages} {2759} (\bibinfo {year} {1992})}\BibitemShut {NoStop}%
\bibitem [{\citenamefont {Miyamoto}\ \emph {et~al.}(1993)\citenamefont
  {Miyamoto}, \citenamefont {Kuga}, \citenamefont {Baba},\ and\ \citenamefont
  {Matsuoka}}]{MiyamotoOL1993}%
  \BibitemOpen
  \bibfield  {author} {\bibinfo {author} {\bibfnamefont {Y.}~\bibnamefont
  {Miyamoto}}, \bibinfo {author} {\bibfnamefont {T.}~\bibnamefont {Kuga}},
  \bibinfo {author} {\bibfnamefont {M.}~\bibnamefont {Baba}},\ and\ \bibinfo
  {author} {\bibfnamefont {M.}~\bibnamefont {Matsuoka}},\ }\bibfield  {title}
  {\bibinfo {title} {Measurement of ultrafast optical pulses with two-photon
  interference},\ }\href {https://doi.org/10.1364/OL.18.000900} {\bibfield
  {journal} {\bibinfo  {journal} {Opt. Lett.}\ }\textbf {\bibinfo {volume}
  {18}},\ \bibinfo {pages} {900} (\bibinfo {year} {1993})}\BibitemShut
  {NoStop}%
\bibitem [{\citenamefont {Yabashi}\ \emph {et~al.}(2002)\citenamefont
  {Yabashi}, \citenamefont {Tamasaku},\ and\ \citenamefont
  {Ishikawa}}]{YabashiPRL2002}%
  \BibitemOpen
  \bibfield  {author} {\bibinfo {author} {\bibfnamefont {M.}~\bibnamefont
  {Yabashi}}, \bibinfo {author} {\bibfnamefont {K.}~\bibnamefont {Tamasaku}},\
  and\ \bibinfo {author} {\bibfnamefont {T.}~\bibnamefont {Ishikawa}},\
  }\bibfield  {title} {\bibinfo {title} {Measurement of x-ray pulse widths by
  intensity interferometry},\ }\href
  {https://doi.org/10.1103/PhysRevLett.88.244801} {\bibfield  {journal}
  {\bibinfo  {journal} {Phys. Rev. Lett.}\ }\textbf {\bibinfo {volume} {88}},\
  \bibinfo {pages} {244801} (\bibinfo {year} {2002})}\BibitemShut {NoStop}%
\bibitem [{\citenamefont {Kameshima}\ \emph {et~al.}(2014)\citenamefont
  {Kameshima}, \citenamefont {Ono}, \citenamefont {Kudo}, \citenamefont
  {Ozaki}, \citenamefont {Kirihara}, \citenamefont {Kobayashi}, \citenamefont
  {Inubushi}, \citenamefont {Yabashi}, \citenamefont {Horigome}, \citenamefont
  {Holland}, \citenamefont {Holland}, \citenamefont {Burt}, \citenamefont
  {Murao},\ and\ \citenamefont {Hatsui}}]{KameshimaRSI2014}%
  \BibitemOpen
  \bibfield  {author} {\bibinfo {author} {\bibfnamefont {T.}~\bibnamefont
  {Kameshima}}, \bibinfo {author} {\bibfnamefont {S.}~\bibnamefont {Ono}},
  \bibinfo {author} {\bibfnamefont {T.}~\bibnamefont {Kudo}}, \bibinfo {author}
  {\bibfnamefont {K.}~\bibnamefont {Ozaki}}, \bibinfo {author} {\bibfnamefont
  {Y.}~\bibnamefont {Kirihara}}, \bibinfo {author} {\bibfnamefont
  {K.}~\bibnamefont {Kobayashi}}, \bibinfo {author} {\bibfnamefont
  {Y.}~\bibnamefont {Inubushi}}, \bibinfo {author} {\bibfnamefont
  {M.}~\bibnamefont {Yabashi}}, \bibinfo {author} {\bibfnamefont
  {T.}~\bibnamefont {Horigome}}, \bibinfo {author} {\bibfnamefont
  {A.}~\bibnamefont {Holland}}, \bibinfo {author} {\bibfnamefont
  {K.}~\bibnamefont {Holland}}, \bibinfo {author} {\bibfnamefont
  {D.}~\bibnamefont {Burt}}, \bibinfo {author} {\bibfnamefont {H.}~\bibnamefont
  {Murao}},\ and\ \bibinfo {author} {\bibfnamefont {T.}~\bibnamefont
  {Hatsui}},\ }\bibfield  {title} {\bibinfo {title} {Development of an x-ray
  pixel detector with multi-port charge-coupled device for x-ray free-electron
  laser experiments},\ }\href {https://doi.org/10.1063/1.4867668} {\bibfield
  {journal} {\bibinfo  {journal} {Review of Scientific Instruments}\ }\textbf
  {\bibinfo {volume} {85}},\ \bibinfo {pages} {033110} (\bibinfo {year}
  {2014})},\ \Eprint {https://arxiv.org/abs/https://doi.org/10.1063/1.4867668}
  {https://doi.org/10.1063/1.4867668} \BibitemShut {NoStop}%
\bibitem [{\citenamefont {Inoue}\ \emph {et~al.}(2018)\citenamefont {Inoue},
  \citenamefont {Hara}, \citenamefont {Inubushi}, \citenamefont {Tono},
  \citenamefont {Inagaki}, \citenamefont {Katayama}, \citenamefont {Amemiya},
  \citenamefont {Tanaka},\ and\ \citenamefont {Yabashi}}]{InouePRAB2018}%
  \BibitemOpen
  \bibfield  {author} {\bibinfo {author} {\bibfnamefont {I.}~\bibnamefont
  {Inoue}}, \bibinfo {author} {\bibfnamefont {T.}~\bibnamefont {Hara}},
  \bibinfo {author} {\bibfnamefont {Y.}~\bibnamefont {Inubushi}}, \bibinfo
  {author} {\bibfnamefont {K.}~\bibnamefont {Tono}}, \bibinfo {author}
  {\bibfnamefont {T.}~\bibnamefont {Inagaki}}, \bibinfo {author} {\bibfnamefont
  {T.}~\bibnamefont {Katayama}}, \bibinfo {author} {\bibfnamefont
  {Y.}~\bibnamefont {Amemiya}}, \bibinfo {author} {\bibfnamefont
  {H.}~\bibnamefont {Tanaka}},\ and\ \bibinfo {author} {\bibfnamefont
  {M.}~\bibnamefont {Yabashi}},\ }\bibfield  {title} {\bibinfo {title} {X-ray
  hanbury brown-twiss interferometry for determination of ultrashort
  electron-bunch duration},\ }\href
  {https://doi.org/10.1103/PhysRevAccelBeams.21.080704} {\bibfield  {journal}
  {\bibinfo  {journal} {Phys. Rev. Accel. Beams}\ }\textbf {\bibinfo {volume}
  {21}},\ \bibinfo {pages} {080704} (\bibinfo {year} {2018})}\BibitemShut
  {NoStop}%
\bibitem [{\citenamefont {Krause}\ and\ \citenamefont
  {Oliver}(1979)}]{KrauseJPCRD1979}%
  \BibitemOpen
  \bibfield  {author} {\bibinfo {author} {\bibfnamefont {M.~O.}\ \bibnamefont
  {Krause}}\ and\ \bibinfo {author} {\bibfnamefont {J.~H.}\ \bibnamefont
  {Oliver}},\ }\bibfield  {title} {\bibinfo {title} {Natural widths of atomic k
  and l levels, k$\alpha$ x‐ray lines and several kll auger lines},\ }\href
  {https://doi.org/10.1063/1.555595} {\bibfield  {journal} {\bibinfo  {journal}
  {Journal of Physical and Chemical Reference Data}\ }\textbf {\bibinfo
  {volume} {8}},\ \bibinfo {pages} {329} (\bibinfo {year} {1979})},\ \Eprint
  {https://arxiv.org/abs/https://doi.org/10.1063/1.555595}
  {https://doi.org/10.1063/1.555595} \BibitemShut {NoStop}%
\bibitem [{Sup()}]{SupplementalMaterial}%
  \BibitemOpen
  \bibfield  {title} {\bibinfo {title} {See supplemental material at url for
  the measured spectral density and the complex degree of coherence of the ni
  $k\alpha$ fluorescence.},\ }\href@noop {} {\ }\BibitemShut {NoStop}%
\bibitem [{\citenamefont {Matsuyama}\ \emph {et~al.}(2018)\citenamefont
  {Matsuyama}, \citenamefont {Inoue}, \citenamefont {Yamada}, \citenamefont
  {Kim}, \citenamefont {Yumoto}, \citenamefont {Inubushi}, \citenamefont
  {Osaka}, \citenamefont {Inoue}, \citenamefont {Koyama}, \citenamefont {Tono},
  \citenamefont {Ohashi}, \citenamefont {Yabashi}, \citenamefont {Ishikawa},\
  and\ \citenamefont {Yamauchi}}]{MatsuyamaSciRep2018}%
  \BibitemOpen
  \bibfield  {author} {\bibinfo {author} {\bibfnamefont {S.}~\bibnamefont
  {Matsuyama}}, \bibinfo {author} {\bibfnamefont {T.}~\bibnamefont {Inoue}},
  \bibinfo {author} {\bibfnamefont {J.}~\bibnamefont {Yamada}}, \bibinfo
  {author} {\bibfnamefont {J.}~\bibnamefont {Kim}}, \bibinfo {author}
  {\bibfnamefont {H.}~\bibnamefont {Yumoto}}, \bibinfo {author} {\bibfnamefont
  {Y.}~\bibnamefont {Inubushi}}, \bibinfo {author} {\bibfnamefont
  {T.}~\bibnamefont {Osaka}}, \bibinfo {author} {\bibfnamefont
  {I.}~\bibnamefont {Inoue}}, \bibinfo {author} {\bibfnamefont
  {T.}~\bibnamefont {Koyama}}, \bibinfo {author} {\bibfnamefont
  {K.}~\bibnamefont {Tono}}, \bibinfo {author} {\bibfnamefont {H.}~\bibnamefont
  {Ohashi}}, \bibinfo {author} {\bibfnamefont {M.}~\bibnamefont {Yabashi}},
  \bibinfo {author} {\bibfnamefont {T.}~\bibnamefont {Ishikawa}},\ and\
  \bibinfo {author} {\bibfnamefont {K.}~\bibnamefont {Yamauchi}},\ }\bibfield
  {title} {\bibinfo {title} {Nanofocusing of x-ray free-electron laser using
  wavefront-corrected multilayer focusing mirrors},\ }\href
  {https://doi.org/10.1038/s41598-018-35611-0} {\bibfield  {journal} {\bibinfo
  {journal} {Scientific Reports}\ }\textbf {\bibinfo {volume} {8}},\ \bibinfo
  {pages} {17440} (\bibinfo {year} {2018})}\BibitemShut {NoStop}%
\bibitem [{\citenamefont {Inoue}\ \emph {et~al.}(2020)\citenamefont {Inoue},
  \citenamefont {Matsuyama}, \citenamefont {Yamada}, \citenamefont {Nakamura},
  \citenamefont {Osaka}, \citenamefont {Inoue}, \citenamefont {Inubushi},
  \citenamefont {Tono}, \citenamefont {Yumoto}, \citenamefont {Koyama},
  \citenamefont {Ohashi}, \citenamefont {Yabashi}, \citenamefont {Ishikawa},\
  and\ \citenamefont {Yamauchi}}]{InoueJSR2020}%
  \BibitemOpen
  \bibfield  {author} {\bibinfo {author} {\bibfnamefont {T.}~\bibnamefont
  {Inoue}}, \bibinfo {author} {\bibfnamefont {S.}~\bibnamefont {Matsuyama}},
  \bibinfo {author} {\bibfnamefont {J.}~\bibnamefont {Yamada}}, \bibinfo
  {author} {\bibfnamefont {N.}~\bibnamefont {Nakamura}}, \bibinfo {author}
  {\bibfnamefont {T.}~\bibnamefont {Osaka}}, \bibinfo {author} {\bibfnamefont
  {I.}~\bibnamefont {Inoue}}, \bibinfo {author} {\bibfnamefont
  {Y.}~\bibnamefont {Inubushi}}, \bibinfo {author} {\bibfnamefont
  {K.}~\bibnamefont {Tono}}, \bibinfo {author} {\bibfnamefont {H.}~\bibnamefont
  {Yumoto}}, \bibinfo {author} {\bibfnamefont {T.}~\bibnamefont {Koyama}},
  \bibinfo {author} {\bibfnamefont {H.}~\bibnamefont {Ohashi}}, \bibinfo
  {author} {\bibfnamefont {M.}~\bibnamefont {Yabashi}}, \bibinfo {author}
  {\bibfnamefont {T.}~\bibnamefont {Ishikawa}},\ and\ \bibinfo {author}
  {\bibfnamefont {K.}~\bibnamefont {Yamauchi}},\ }\bibfield  {title} {\bibinfo
  {title} {{Generation of an X-ray nanobeam of a free-electron laser using
  reflective optics with speckle interferometry}},\ }\href
  {https://doi.org/10.1107/S1600577520006980} {\bibfield  {journal} {\bibinfo
  {journal} {Journal of Synchrotron Radiation}\ }\textbf {\bibinfo {volume}
  {27}},\ \bibinfo {pages} {883} (\bibinfo {year} {2020})}\BibitemShut
  {NoStop}%
\bibitem [{\citenamefont {Huang}\ \emph {et~al.}(2017)\citenamefont {Huang},
  \citenamefont {Ding}, \citenamefont {Feng}, \citenamefont {Hemsing},
  \citenamefont {Huang}, \citenamefont {Krzywinski}, \citenamefont {Lutman},
  \citenamefont {Marinelli}, \citenamefont {Maxwell},\ and\ \citenamefont
  {Zhu}}]{HuangPRL2017}%
  \BibitemOpen
  \bibfield  {author} {\bibinfo {author} {\bibfnamefont {S.}~\bibnamefont
  {Huang}}, \bibinfo {author} {\bibfnamefont {Y.}~\bibnamefont {Ding}},
  \bibinfo {author} {\bibfnamefont {Y.}~\bibnamefont {Feng}}, \bibinfo {author}
  {\bibfnamefont {E.}~\bibnamefont {Hemsing}}, \bibinfo {author} {\bibfnamefont
  {Z.}~\bibnamefont {Huang}}, \bibinfo {author} {\bibfnamefont
  {J.}~\bibnamefont {Krzywinski}}, \bibinfo {author} {\bibfnamefont {A.~A.}\
  \bibnamefont {Lutman}}, \bibinfo {author} {\bibfnamefont {A.}~\bibnamefont
  {Marinelli}}, \bibinfo {author} {\bibfnamefont {T.~J.}\ \bibnamefont
  {Maxwell}},\ and\ \bibinfo {author} {\bibfnamefont {D.}~\bibnamefont {Zhu}},\
  }\bibfield  {title} {\bibinfo {title} {Generating single-spike hard x-ray
  pulses with nonlinear bunch compression in free-electron lasers},\ }\href
  {https://doi.org/10.1103/PhysRevLett.119.154801} {\bibfield  {journal}
  {\bibinfo  {journal} {Phys. Rev. Lett.}\ }\textbf {\bibinfo {volume} {119}},\
  \bibinfo {pages} {154801} (\bibinfo {year} {2017})}\BibitemShut {NoStop}%
\bibitem [{\citenamefont {Marinelli}\ \emph {et~al.}(2017)\citenamefont
  {Marinelli}, \citenamefont {MacArthur}, \citenamefont {Emma}, \citenamefont
  {Guetg}, \citenamefont {Field}, \citenamefont {Kharakh}, \citenamefont
  {Lutman}, \citenamefont {Ding},\ and\ \citenamefont
  {Huang}}]{MarinelliAPL2017}%
  \BibitemOpen
  \bibfield  {author} {\bibinfo {author} {\bibfnamefont {A.}~\bibnamefont
  {Marinelli}}, \bibinfo {author} {\bibfnamefont {J.}~\bibnamefont
  {MacArthur}}, \bibinfo {author} {\bibfnamefont {P.}~\bibnamefont {Emma}},
  \bibinfo {author} {\bibfnamefont {M.}~\bibnamefont {Guetg}}, \bibinfo
  {author} {\bibfnamefont {C.}~\bibnamefont {Field}}, \bibinfo {author}
  {\bibfnamefont {D.}~\bibnamefont {Kharakh}}, \bibinfo {author} {\bibfnamefont
  {A.~A.}\ \bibnamefont {Lutman}}, \bibinfo {author} {\bibfnamefont
  {Y.}~\bibnamefont {Ding}},\ and\ \bibinfo {author} {\bibfnamefont
  {Z.}~\bibnamefont {Huang}},\ }\bibfield  {title} {\bibinfo {title}
  {Experimental demonstration of a single-spike hard-x-ray free-electron laser
  starting from noise},\ }\href {https://doi.org/10.1063/1.4990716} {\bibfield
  {journal} {\bibinfo  {journal} {Applied Physics Letters}\ }\textbf {\bibinfo
  {volume} {111}},\ \bibinfo {pages} {151101} (\bibinfo {year} {2017})},\
  \Eprint {https://arxiv.org/abs/https://doi.org/10.1063/1.4990716}
  {https://doi.org/10.1063/1.4990716} \BibitemShut {NoStop}%
\end{thebibliography}%

\end{document}